\documentclass[10pt, conference, compsocconf]{IEEEtran}   
\usepackage{flushend}
\usepackage{graphicx}				
\graphicspath{{plots/}}							

\usepackage{etoolbox}

\usepackage{color}

\usepackage{amssymb, amsmath}
\usepackage{mathtools}
\usepackage{booktabs}
\usepackage{todonotes}
\usepackage{paralist}
\usepackage{verbatim}

\usepackage{clrs+} 
\usepackage[noend]{algpseudocode}
\usepackage[algo2e,ruled,vlined,linesnumbered]{algorithm2e}
\usepackage{varwidth}
\usepackage{multirow}
\usepackage{rotating, xcolor}
\usepackage{subfig}
\usepackage{enumitem}
\usepackage[hyphens]{url}
\usepackage{float}

\newcommand{\qc}{\c{c}}

\newcommand{\minusone}{\text{-}1}

\newcommand{\mA}{\mathbf{A}}

\newcommand{\transpose}     {^{\mbox{\scriptsize \sf T}}}

\newcommand{\mP}{\mathbf{P}}

\newcommand{\dnnz}{\id{nnz}}

\newcommand{\dth}{th}

\IEEEoverridecommandlockouts

\title{The Reverse Cuthill-McKee Algorithm in Distributed-Memory \thanks{A two-page abstract of this work has appeared at the SIAM Workshop on Combinatorial 
Scientific Computing. Short abstracts on SIAM conferences are not archived and does not appear in published proceedings.}}

\author{Ariful~Azad, Mathias Jacquelin,
Ayd\i n Bulu\qc,  Esmond G. Ng \\
  {\small E-mail: \{azad, mjacquelin, abuluc,  egng\}@lbl.gov }\\ 
  Computational Research Division \\
  Lawrence Berkeley National Laboratory \\
}

\date{}	

\begin{document}
\maketitle

\begin{abstract}

Ordering vertices of a graph is key to minimize fill-in and data structure size in sparse direct solvers, maximize locality in iterative solvers, and improve performance in graph algorithms. Except for naturally parallelizable ordering methods such as nested dissection, many important ordering methods have not been efficiently mapped to distributed-memory architectures. 
In this paper, we present the first-ever distributed-memory implementation of the reverse Cuthill-McKee (RCM) algorithm for reducing the profile of a sparse matrix. 
Our parallelization uses a two-dimensional sparse matrix decomposition. We achieve high performance by decomposing the problem into a small number of primitives and 
utilizing optimized implementations of these primitives.
Our implementation shows strong scaling up to 1024 cores for smaller matrices and up to 4096 cores for larger matrices. 
\end{abstract}

\section{Introduction}
Reordering a sparse matrix to reduce its bandwidth or profile can speed up many sparse matrix computations~\cite{duff1989use, duff1989effect}.
For example, a matrix with a small profile is useful in direct methods for solving sparse linear systems since it allows a simple data structure to be used.  
It is also useful in iterative methods because the nonzero elements will be clustered close to the diagonal, thereby enhancing data locality.
Given a symmetric matrix $\mA$, a bandwidth-reduction ordering aims to find a permutation $\mP$ so that the bandwidth of $\mP\mA\mP\transpose$  is small.
Since obtaining a reordering to minimize bandwidth is an NP-complete problem~\cite{papadimitriou1976np}, various heuristics are used in practice such as Cuthill-McKee, Reverse Cuthill-McKee (RCM), and Sloan's algorithms~\cite{cuthill1969reducing, GeorgeLiu81, sloan1986algorithm}. 
This paper solely focuses on the RCM algorithm~\cite{GeorgeLiu81} because, with careful algorithm design, it is amenable to massive distributed-memory parallelism -- the primary topic of interest of this paper.  

\begin{figure}[!t]
   \centering
   
   \includegraphics[scale=.5]{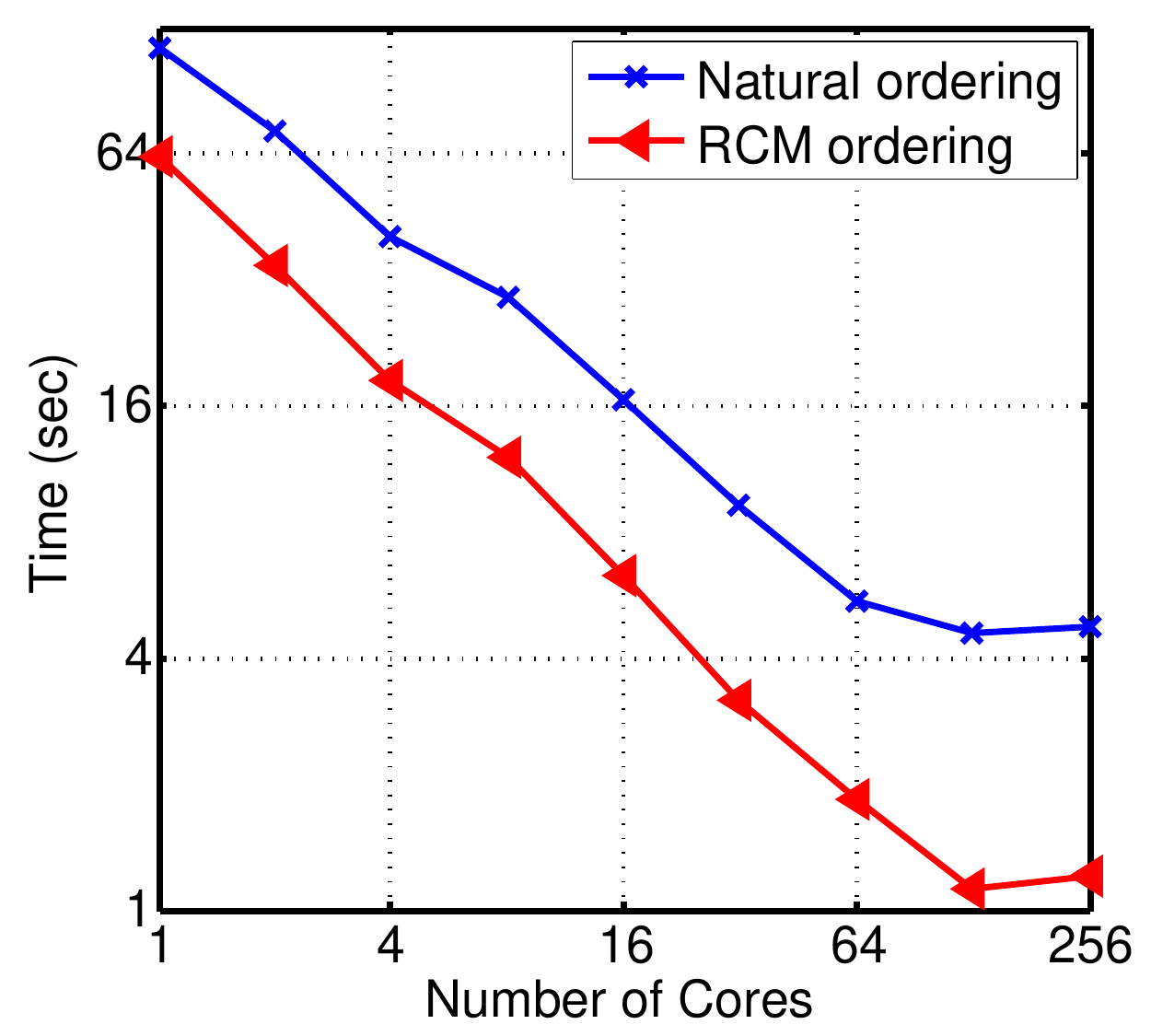}
\caption{The time to solve \texttt{thermal2} using the conjugate gradient method with block Jacobi preconditioner in PETSc on NERSC/Edison. \texttt{thermal2} has 1.2M rows and columns and 4.9M nonzeros. The bandwidths of the original and RCM-permuted matrix are 1,226,000 and 795, respectively.}
   \label{fig:gathertime}
   \vspace{-6pt}
\end{figure}



The need for distributed memory RCM algorithms is driven by the move to extreme-scale computing.
In a large scale scientific application, the matrix has often been distributed already by the time one arrives at the numerical phases of sparse matrix computations.
Hence, it would be a waste of time to gather a distributed matrix onto a single processor to compute an ordering using sequential or multithreaded algorithms. Furthermore, by clustering the nonzeros closer to the diagonal, RCM ordering not only increases cache performance of iterative solvers, but it can often restrict the communication to resemble more of a nearest-neighbor pattern. Figure~\ref{fig:gathertime} illustrates the performance effects of RCM ordering on a preconditioned conjugate gradient solver of the popular PETSc package~\cite{petsc-user-ref}. Notice that the performance benefit of RCM ordering actually
increases as the number of cores increases, possible due to reduced communication costs.
Our goal in this paper is therefore to design and develop a scalable distributed-memory parallel implementation of the RCM algorithm~\cite{GeorgeLiu81}.

Similar to many sparse matrix computations, RCM ordering has been shown to be a difficult problem to parallelize~\cite{spmp-rcm}. The RCM algorithm relies on the repeated application of breadth-first search (BFS) and its computational load is highly dynamic, especially if the graph has high diameter. 
The problem exacerbates on high concurrency, where load imbalance and communication overhead degrade the performance of the parallel algorithm.
RCM is harder to parallelize than BFS primarily for two reasons. First,
RCM imposes additional restrictions on how the vertices are numbered
in the graph traversal, limiting available parallelism. Second, RCM is often used for matrices with medium-to-high diameter while most of the existing work on parallel BFS has focused on low-diameter graphs such as synthetic
graphs used by the Graph500 benchmark. A higher diameter increases the 
critical path of level-synchronous BFS algorithms, also limiting available parallelism.



In this paper, we aim to overcome these challenges by using the graph-matrix duality and replacing unstructured graph operations by structured matrix/vector operations.
We make the following contributions:
\begin{itemize}
\item We present a scalable  distributed-memory algorithm for RCM ordering. Our algorithm relies on a handful of bulk-synchronous parallel primitives that are optimized for both shared-memory and massively parallel distributed memory systems. 
\item The quality (bandwidth and envelope) of ordering from our distributed-memory implementation is comparable to the state-of-the-art and remains insensitive to the degree of concurrency. 
\item We provide a hybrid OpenMP-MPI implementation of the RCM ordering that attains up to 38$\times$ speedup on matrices from various applications on 1024 cores of a Cray XC30 supercomputer. 
We provide detailed performance evaluation on up to 4096 cores, which sheds light on the performance bottlenecks and opportunities for future research. 
\end{itemize}


\section{Preliminaries}
\label{sec:background}

\subsection{Serial Algorithms}


Let $n$ be the number of columns in a symmetric matrix $\mA$.
Let $f_i(\mA)$ be the the column subscript of the first nonzero element in column $i$ of $\mA$: $f_i(\mA) = \min\{j ~~|~ a_{ij} \neq 0\}$.
The \emph{$i$-th bandwidth} $\beta_i(\mA)$ is defined as 
$\beta_i(\mA) = i - f_i(\mA)$. The overall 
\emph{bandwidth} of matrix $\mA$ is denoted $\beta(\mA)$ and is defined as $\beta(\mA) = \max\{\beta_i(\mA) ~~|~ 1 \leq i \leq n\}$.
Using these notations, the envelope $Env(\mA)$ is defined
as:
\begin{equation*}
\vspace{-1mm}
Env(\mA) = \left\{\left\{i, j\right\} ~~|~ 0 < j - i \leq \beta_i(\mA) , 1 \leq i,j \leq n \right\}.
\end{equation*}
\vspace{-1mm}
The quantity $|Env(\mA)|$ is called the profile or envelope size of $\mA$.

Reverse Cuthill-McKee
ordering, or RCM, introduced by George~\cite{george1971}, is a variant of the Cuthill-McKee ordering~\cite{cuthill1969reducing}, 
which aims at reducing the \emph{bandwidth} of a sparse symmetric matrix $\mA$. This is of particular
importance when the matrix is to be stored using a profile based format.
%
Finding a reordering of the rows/columns of $\mA$ corresponds to the process of 
labeling vertices of the graph $\mathcal{G}(\mA)$ associated
with $\mA$.

RCM repeatedly labels vertices
adjacent to the current vertex $v_i$ until all have been labeled, as depicted in 
Algorithm~\ref{algo:seq_rcm}. The algorithm essentially processes vertices levels by levels
and labels them in reverse order. The resulting reordered matrix often has a smaller profile~\cite{george1971}.

\begin{algorithm2e}
\DontPrintSemicolon

$v_1 \gets r$\;
\For{$i = 1$ to $n$}{
  Find all the unnumbered neighbors of the vertex $v_i$\;
  Label the vertices found in increasing order of degree\;
}
The Reverse Cuthill-McKee ordering is given by $w_1,w_2,\ldots,w_n$
where $w_i = v_{n-i+1}$ for $i=1,\ldots,n$.

\label{algo:serialRCM}
\caption{Reverse Cuthill-McKee algorithm\label{algo:seq_rcm}}
\end{algorithm2e}

For the graph $\mathcal{G}(\mA) = (V,E)$ associated with $\mA$, $V$ is the set of vertices and $E$ is the set of edges in $\mathcal{G}(\mA)$.
The number of vertices in $\mathcal{G}(\mA)$ is denoted $n = |V|$ and $m$ is the number of edges $m = |E|$.
The \emph{eccentricity}~\cite{berge} of a vertex $v$ is defined as
$  \ell(v) = \max\{d(v,w)~~|~w \in V \},$
where $d(v,w)$ is the graph distance between $v$ and $w$.

The \emph{rooted level structure}~\cite{arany} of a vertex $v \in V$ is
the \emph{partioning} $\mathcal{L}(v)$ of $V$ satisfying
\begin{equation*}
  \mathcal{L}(v) = \{L_0(v), L_1(v),\ldots,L_{\ell(v)}(v)\},
\vspace*{-2mm}
\end{equation*}
where
\vspace*{-2mm}
\begin{equation*}
  L_0(v) = \{v\}, \quad L_1(v) = \textnormal{Adj}(L_0(v)),
\end{equation*}
\begin{equation*}
  L_i(v) = \textnormal{Adj}(L_{i-1}(v)) \setminus L_{i-2}(v), \quad i= 2,3,\ldots,\ell(v).
\end{equation*}

The \emph{length} of $\mathcal{L}(v)$ corresponds to the eccentricity $\ell(v)$
while the \emph{width} $\nu(v)$ of $\mathcal{L}(v)$ is defined by
\begin{equation*}
  \nu(v) = \max\{|L_i(v)|~~|~0 \leq i \leq \ell(v)\}.
\end{equation*}

The first vertex being labeled strongly impacts the bandwidth of the permuted 
matrix. Experience shows that it is better to start with a node having a large eccentricity. 
A \emph{peripheral} vertex is a vertex with maximum eccentricity. Finding such vertex is prohibitively expensive, and a 
common heuristic is to use a \emph{pseudo-peripheral vertex} instead~\cite{george1979,GeorgeLiu81}.
A \emph{pseudo-peripheral vertex} is a vertex displaying a 
high eccentricity, as close to the graph diameter as 
possible. Gibbs et al.~\cite{gibbs1976algorithm} introduced an algorithm to find such a 
vertex, which is later refined by George and Liu~\cite{george1979}. The process, given in 
Algorithm~\ref{algo:seq_pseudo}, starts with an arbitrary 
vertex in $V$ and computes its rooted level structure. Then, a 
vertex in the last level is picked and the corresponding
level structure is computed. This process is repeated until
the number of levels in the rooted level structure 
converges.
Computing the level structure corresponds to a BFS of the graph $\mathcal{G}(\mA)$.

\begin{algorithm2e}
\DontPrintSemicolon

$r \gets \{\textnormal{arbitrary vertex in } V\}$ \;
$\mathcal{L}(r) \gets \{ L_0(r), L_1(r),\ldots,L_{\ell(r)}(r)\}$ \;
$\textnormal{nlvl} \gets \ell(r) - 1$ \;

\While{$\ell(r) > \textnormal{nlvl}$}{
  $\textnormal{nlvl} \gets \ell(r)$\;
(Shrink last level): Choose a vertex $v$ in $L_{\ell(r)}(r)$ of minimum degree.\;
  $\mathcal{L}(v) \gets \{ L_0(v), L_1(v),\ldots,L_{\ell(v)}(v)\}$\;
  $r \gets v$\;
}
\caption{An algorithm to find a pseudo-peripheral vertex $r$\label{algo:seq_pseudo}}
\end{algorithm2e}

\begin{table*}[!t]
   \centering
    \caption{Matrix-algebraic primitives needed for the RCM algorithm. }
    \scalebox{1}{
   \begin{tabular}{@{} llllcc @{}} 
      \toprule
      \multirow{2}{*}{Function} & \multirow{2}{*}{Arguments} & \multirow{2}{*}{Returns}  & Serial &  \multirow{2}{*}{Communication} \\
      &  &   & Complexity & \\
      \toprule

       \multirow{2}{*}{\Call{Ind}{}} & \multirow{2}{*}{$x$: a sparse vector} &  local indices of  &  \multirow{2}{*}{$O(\dnnz(x))$}  & None\\
       &  &  nonzero entries of $x$ &  &   & \\
      \midrule
          
     \multirow{4}{*}{\Call{Select}{}} & $x$: a sparse vector & $z \gets$ an empty sparse vector &    & \\
       & $y$: a dense vector & for $i \in \Call{Ind}{x}$   &  $O(\dnnz(x))$  & None\\
       &  $\id{expr}$: logical expr.\ on $y$  &  \ \ \ {\bf if} $(\id{expr}(y[i]))$ {\bf then}&   & \\
       &  assume $\id{size}(x) = \id{size}(y)$ &  \ \ \ \ \ \ $z[i]\gets x[i]$ &   & \\
  \midrule

 \multirow{2}{*}{\Call{Set}{}} & $x$: a sparse vector & for $i \in \Call{Ind}{x}$    & $O(\id{nnz}(x))$ & None\\
       &  $y$: a dense vector & \ \ \  $y[i] \gets x[i]$ &   & \\
       \midrule
       
	\multirow{3}{*}{\Call{SpMSpV}{}} & $\mA$: a sparse matrix &   &   \multirow{3}{*}{$\sum\limits_{\mathclap{k \in \Call{Ind}{x}}} \id{nnz}(\mA(:,k))$ }  &  AllGather \& \\
     & $x$: a sparse vector  & returns $\mA \cdot x$   &       & AlltoAll on\\
       & SR: a semiring & &     & subcommunicator~\cite{bfs:11} \\
       \midrule
       \multirow{3}{*}{\Call{Reduce}{}} & $x$: a sparse vector &  mv = maximum value in $y$ &   \multirow{3}{*}{$O(\dnnz(x))$ }  & \\
     & $y$: a dense vector  &   for $i \in \Call{Ind}{x}$  &       &   AllReduce\\
       & a reduction operation & \ \ \  mv $\gets$ min\{mv, $y[i]$\} &     & \\
       
       \midrule
       \multirow{4}{*}{\Call{SortPerm}{}} &  &  $T \gets$ an empty array of tuples &    & \multirow{4}{*}{AllToAll}\\
     &  $x$: a sparse vector &   for $i \in \Call{Ind}{x}$  &   $O(\dnnz(x) \, \log{\dnnz(x)})$     &   \\
       & $y$: a dense vector & \ \ \  $T[i]$ $\gets (x[i], y[i], i)$ &      & \\
       &  & sort $T$ and return the permutation &     & \\
       \toprule
   \end{tabular}}
   \label{tab:operations}
\end{table*}

\subsection{Previous Work on Parallel RCM}
There is a strong connection between BFS, finding a pseudo-peripheral vertex and computing the RCM ordering. However, RCM is often used
for matrices with higher diameter than graphs for which parallel BFS is often optimized~\cite{bfs:11,checconi2012breaking}. In addition, computing the actual RCM ordering is even harder to parallelize because the vertices within each level of the traversal tree has to be ordered by degree. 

Computing sparse matrix orderings in parallel have received intermittent attention over the last few decades. 
While RCM is often used to accelerate iterative solvers, one of its first reported supercomputer-scale 
implementations was in the context of direct solvers by Ashcraft et al.~\cite{ashcraft1987progress} on a CRAY X-MP.
It is hard to compare results from 30 years ago, where both the architectures and the sizes of matrices were significantly different. 
Karantasis et al.~\cite{spmp-rcm} recently studied the shared-memory parallelization of various reordering algorithms including RCM.

\section{Algorithms based on matrix algebra}
\label{sec:dist-algo}

In this section, we present a distributed memory algorithm for computing the RCM ordering that takes advantage of the equivalence between graph algorithms and sparse matrix algebra to exploit latest hardware platforms. The $\dnnz()$ function computes the number of nonzeros in its input, e.g.,  $\dnnz(\mathbf{x})$ returns the number of nonzeros in  $\mathbf{x}$. We utilize
the Matlab colon notation: $\mA(:,i)$ denotes the $i$\dth\ column, and $\mA(i,:)$ denotes  the $i$\dth\ row.

\begin{algorithm2e*}
\DontPrintSemicolon

	$\mathcal{R} \gets -1$\ \tcp*{Dense vector storing the ordering of all vertices; initialized to -1}  
	$\mathcal{L}_{cur} \gets \{r\}$ , $\mathcal{L}_{next} \gets \phi$ \tcp*{Current and next BFS levels (frontier)}
	$\mathcal{R}[r] \gets 0$ \tcp*{label of $r$ is set to 0}
    $\textnormal{nv} \gets 1$ \tcp*{Number of vertices labeled so far}
 	\While{$\mathcal{L}_{cur} \neq \phi$}{ \label{algo:li_while1}
    	$\mathcal{L}_{cur} \gets \Call{Set}{\mathcal{L}_{cur}, \mathcal{R}}$ \; 
 		$\mathcal{L}_{next} \gets$ \Call{SpMSpV}{$\mA,\mathcal{L}_{cur} $, SR=(select2nd, min)}  \tcp*{Visit neighbors of the frontier} 
		$\mathcal{L}_{next} \gets \Call{Select}{\mathcal{L}_{next} , \mathcal{R}= \minusone}$  \tcp*{Keep unvisited vertices} 

		$\mathcal{R}_{next} \gets \Call{SortPerm}{\mathcal{L}_{next}, \mathcal{D}}$ \tcp*{Lexicographically sorted permutation of vertices in the next frontier based on the (parent\_order, degree) pair}
        $\mathcal{R}_{next} \gets \mathcal{R}_{next} + \textnormal{nv}$ \tcp*{Global ordering}
        $\textnormal{nv} \gets \textnormal{nv} + \id{nnz}({R}_{next})$\;
		$\mathcal{R} \gets \Call{Set}{\mathcal{R}, \mathcal{R}_{next}  }$ \tcp*{Set orders of the newly visited vertices}
		$\mathcal{L}_{cur}  \gets \mathcal{L}_{next} $ \; \label{algo:li_while2}	
 	} 
 
 \Return{$\mathcal{R}$ \textnormal{in the reverse order}}
\caption{Reverse Cuthill-McKee algorithm in terms of matrix-algebraic operations. {\bf Inputs:} a sparse adjacency matrix $\mA$, a dense vector $\mathcal{D}$ storing the degrees of all vertices, and a pseudo-peripheral vertex $r$. {\bf Output:} the RCM ordering $\mathcal{R}$. \label{algo:dist_RCM}}
\end{algorithm2e*}

\SetKwRepeat{Do}{do}{while}
\begin{algorithm2e*}
\DontPrintSemicolon

$r \gets \{\textnormal{arbitrary vertex in } V\}$ \;
 $\ell \gets $ 0 , $\textnormal{nlvl} \gets \ell - 1$ \tcp*{Number of levels in current and previous BFS trees}

\While{$\ell > \textnormal{nlvl}$}
 {
	$\mathcal{L} \gets -1$\ \tcp*{Dense vector storing the BFS level that each vertex belongs to}   
		$\mathcal{L}_{cur} \gets \{r\}$ , $\mathcal{L}_{next} \gets \phi$ \tcp*{Current and next BFS levels (frontier)}
	$\textnormal{nlvl} \gets \ell$ , $\mathcal{L}[r] \gets 0$ \tcp*{Level of $r$ is set to 0}
 	\Do{$\mathcal{L}_{next} \neq \phi$}{
    	$\mathcal{L}_{cur} \gets \Call{Set}{\mathcal{L}_{cur}, \mathcal{L}}$ \;
 		$\mathcal{L}_{next} \gets$ \Call{SpMSpV}{$\mA,\mathcal{L}_{cur} $, SR=(select2nd, min)}  \tcp*{Visit neighbors} 
		$\mathcal{L}_{next} \gets \Call{Select}{\mathcal{L}_{next} , \mathcal{L}= \minusone}$  \tcp*{Keep unvisited vertices} 
		\If{$\mathcal{L}_{next} \neq \phi$}{
			$\mathcal{L} \gets \Call{Set}{\mathcal{L}, \mathcal{L}_{next}  }$ \tcp*{Set levels of newly visited vertices}
			$\mathcal{L}_{cur}  \gets \mathcal{L}_{next} $ 
			
		}
		$\ell \gets \ell + 1$ \;
		
 	}
	$r \gets \Call{Reduce}{\mathcal{L}_{cur}, \mathcal{D}}$  \tcp*{Find the vertex with the minimum degree} 
 }

\caption{Algorithm to find a pseudo-peripheral vertex $r$ in terms of matrix-algebraic operations. {\bf Inputs:} a sparse adjacency matrix $\mA$ and a dense vector $\mathcal{D}$ storing the degrees of all vertices. {\bf Output:} a pseudo-peripheral vertex $r$. \label{algo:dist_pseudo}}
\end{algorithm2e*}

\subsection{Primitives for the RCM algorithm}
With an aim to design a scalable parallel algorithm, we redesign the sequential algorithms for RCM ordering  (Algorithm~\ref{algo:seq_rcm}) and finding a pseudo-peripheral vertex (Algorithm~\ref{algo:seq_pseudo}) in terms of matrix-algebraic operations. 
For this purpose, we use the primitives summarized in Table~\ref{tab:operations}.
A sparse vector is used to represent a subset of vertices.
Consider that a subset of $n_1$ unique vertices $V_1$ is represented by a sparse vector $x$. Then, for each vertex $v_i$ in $V_1$, there is a nonzero entry in the $i$-th location of $x$ (i.e., $x[i]\neq 0$), and the number of nonzeros $\dnnz(x)$ in $x$ is equal to $n_1$.
The nonzero entries of the sparse vector can store arbitrary values depending on the context of the algorithm.
By contrast, a dense vector $y$ stores information about all vertices in the graph, hence the length of $y$ is always $n$. 
In sparse matrix computations, we often are interested in where the nonzero elements are in the sparse matrices.  Hence, if $\mA$ is an $n \times n$ sparse symmetric matrix, then we use a graph $G=(V,E)$ to describe the sparsity structure of $\mA$: $V = \{ v_1 , v_2 , \cdots , v_n \}$, where $v_i$ corresponds to row/column $i$ of $\mA$, for $1 \leq i \leq n$, and $\{ v_i , v_j \} \in E$ when $\mA_{i,j} \neq 0$.

Let $x$ be a sparse vector, $y$ be a dense vector, $\id{expr}$ be a logical unary operation and $\mA$ be a sparse matrix.
\Call{Ind}{$x$} returns the indices of nonzero entries in $x$.
\Call{Select}{$x,y, \id{expr}$} returns every nonzero entry $x[i]$ of $x$ where $\id{expr}(y[i])$ is true.
\Call{Set}{$y,x$} replaces values of $y$ by corresponding nonzero entries of $x$ (other entries of $y$ remain unchanged).
\Call{Reduce}{$x,y,op$} considers only the nonzero indices $I$ of the sparse vector $x$ and performs a reduction on the values of the dense vector $y[I]$ using the operation $op$. In Table~\ref{tab:operations}, we provide an example where the reduction operation is to find the minimum value. 
\Call{SortPerm}{$x,y$} creates a tuple $(x[i], y[i], i)$ for each nonzero index $i$ in $x$. The list of tuples are lexicographically sorted and the function returns the sorted permutation (the indices passed to the sorting routine are used to obtain the permutation).
Finally, \Call{SpMSpv}{$\mA, x, \id{SR}$} performs a sparse matrix-sparse vector multiplication over the semiring $\id{SR}$.  
For the purposes of this work, a semiring is defined over (potentially separate) sets of `scalars', and has its two 
operations `multiplication' and `addition' redefined. We refer to a semiring by listing its scaling operations, such as the \emph{(multiply, add)} semiring.
The usual semiring multiply for BFS is \emph{select2nd}, which returns the second value it is passed. 
The BFS semiring is defined over two sets: the matrix elements are from the set of binary numbers, whereas the vector elements are from the set of integers. 

\begin{figure*}[htbp]
   \centering
   \includegraphics[scale=.5]{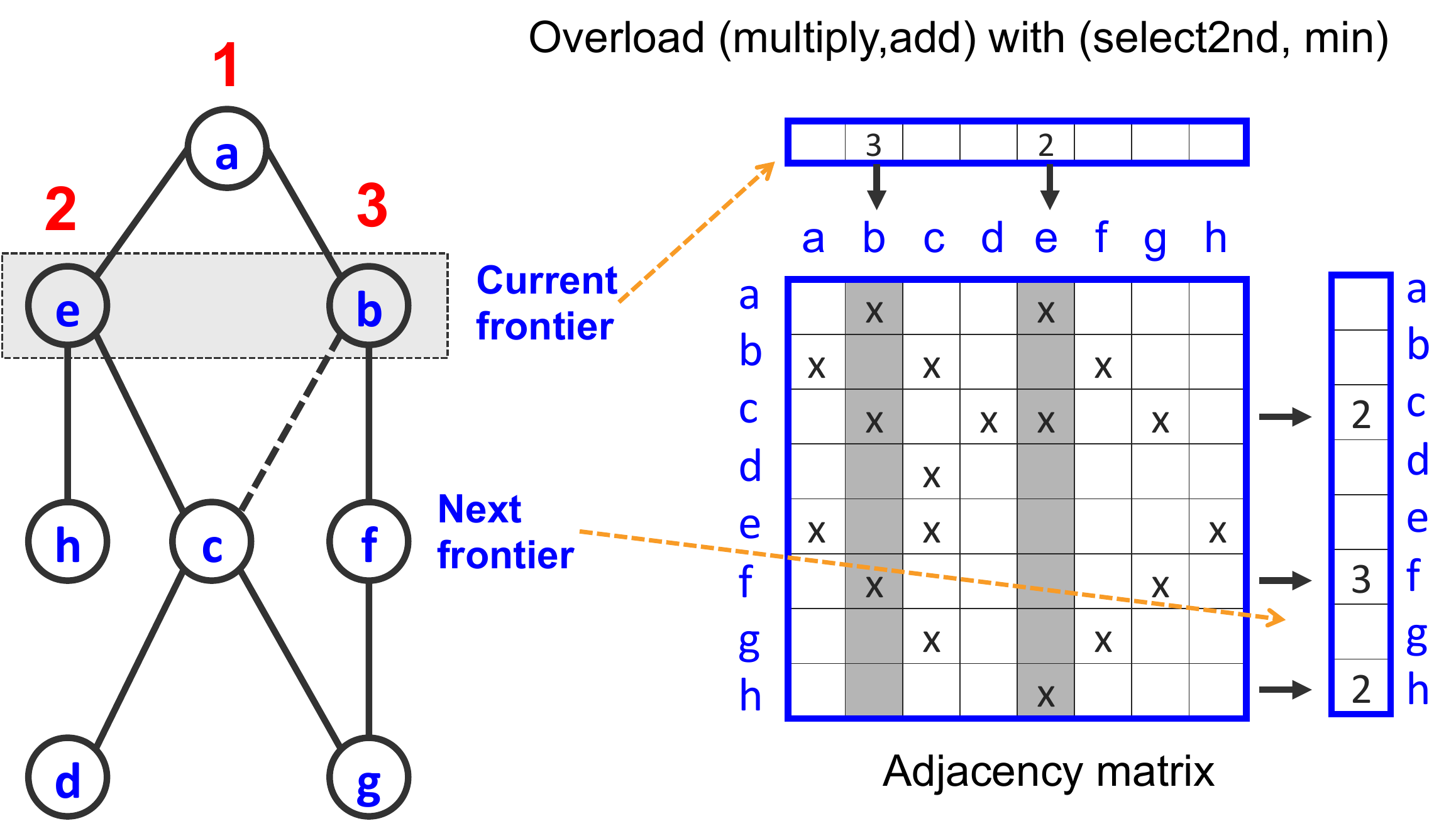} 
   \caption{An example of using specialized sparse matrix-sparse vector multiplication in exploring the next-level vertices in the RCM algorithm. On the left, we show a BFS tree rooted at vertex $a$. The current frontier is the set of two vertices \{$e$, $b$\}. Labels of the already explored vertices are shown in red numbers. The adjacency matrix representing the graph is shown on the right. The input sparse vector corresponding to the current frontier is show at the top of the matrix. The sparse vector has two nonzero entries corresponding to $e$ and  $b$. The values of the nonzero entries are the labels of the vertices.The SpMSpV algorithm over a \emph{(select2nd, min)} semiring then selects the columns of the matrix corresponding to the nonzero indices of the vector shown in gray columns and 
 retains the minimum product in each row where there exists at least one nonzero entry in the selected columns. The indices of the output vector represent the vertices in the next level while the values in the vector denote the labels of parent vertices. The \emph{(select2nd, min)} semiring ensures that a child $v$ always attaches itself to a parent with the minimum label among all of $v$'s visited neighbors.
 Therefore, for vertex $c$, vertex $e$ is selected as a parent instead of vertex $b$ because the former has a smaller label than the latter.}
   \label{fig:example}
\end{figure*}

\subsection{The RCM algorithm using matrix-algebraic primitives} \label{sec:primitives}
Algorithm~\ref{algo:dist_RCM} describes the RCM algorithm using the operations from Table~\ref{tab:operations}.  
Here, we assume that the graph is connected. The case for more than connected components can be handled by repeatedly invoking Algorithm~\ref{algo:dist_RCM} for each connected component.
Algorithm~\ref{algo:dist_RCM} takes a sparse adjacency matrix $\mA$ and a pseudo-peripheral vertex $r$ as inputs, and returns the RCM ordering as a dense vector $\mathcal{R}$. 

The $i$-th element $\mathcal{R}[i]$ of $\mathcal{R}$ is initialized to $-1$ and it retains the initial value until the $i$-th vertex is visited and labeled.  
Let $\mathcal{L}_{cur}$ and $\mathcal{L}_{next}$ be the set of vertices in the current and the next BFS level, respectively. 
$\mathcal{L}_{cur}$ is called the BFS frontier (the set of current active vertices).
The algorithm starts by labeling the pseudo-peripheral vertex $r$, inserting it into $\mathcal{L}_{cur}$.

The while loop (lines \ref{algo:li_while1}-\ref{algo:li_while2} of Algorithm~\ref{algo:dist_RCM}) explores the vertices level-by-level until the frontier $\mathcal{L}_{cur}$ becomes empty.
Vertices in $\mathcal{L}_{cur}$ have been already labeled  in the previous iteration or during initialization. 
Hence each iteration of the while loop traverses the unvisited neighbors $\mathcal{L}_{next}$ of $\mathcal{L}_{cur}$ and labels vertices in $\mathcal{L}_{next}$.

At the beginning of the while loop of Algorithm~\ref{algo:dist_RCM}, the values of the sparse vector $\mathcal{L}_{cur}$ are set to the labels of vertices.
Next, we discover vertices $\mathcal{L}_{next}$ that are adjacent to the current frontier $\mathcal{L}_{cur}$ by multiplying $\mA$ by $\mathcal{L}_{cur}$ over (\emph{select2nd, min}) semiring (line 7).
Here the overloaded multiplication operation \emph{selec2nd} passes the labels of parents to the children and the overloaded addition operation \emph{min} ensures that each vertex in $\mathcal{L}_{next}$ becomes a child of a vertex in $\mathcal{L}_{cur}$ with the minimum label.
This step is explained in Figure~\ref{fig:example} with an example. 
Notice that the specific operator overloading by a (\emph{select2nd, min}) semiring makes the vertex exploration deterministic, and parallelizing  the specialized SpMSpV becomes more challenging than traditional BFS.
Thankfully, the linear algebraic formulation with its operator overloading feature hides the RCM-specific complexity and is able to achieve high performance despite the deterministic nature of the computation.

After the vertices in the next level are discovered via SpMSpV, previously labeled vertices are removed from $\mathcal{L}_{next}$ (line 8).
The next step is to label the vertices in $\mathcal{L}_{next}$. 
For this purpose, we sort vertices in $\mathcal{L}_{next}$ based on the labels of the parents of $\mathcal{L}_{next}$ and the degrees of vertices $\mathcal{L}_{next}$. 
The sorted permutation of vertices in $\mathcal{L}_{next}$ is used to set their labels.
Finally, we update the labels of the newly visited vertices (line 12) and proceed to the next iteration of the while loop.
When the current frontier becomes empty, Algorithm~\ref{algo:dist_RCM} returns with the RCM ordering by reversing the labels of the vertices.

\subsection{Finding the pseudo-peripheral vertex} \label{sec:dist-pseudo}

Algorithm~\ref{algo:dist_pseudo} finds a pseudo-peripheral vertex using matrix-algebraic operations from Table~\ref{tab:operations}.
Similar to Algorithm~\ref{algo:seq_pseudo}, Algorithm~\ref{algo:dist_pseudo} starts with an arbitrary vertex $r$.
Initially, the number of levels in the current and previous BFSs are set to $0$ and $-1$, respectively, so that the while loop in line 4 iterates at least twice before termination.
Each iteration of the while loop in line 4 runs a full BFS starting with $r$. 
As before, $\mathcal{L}_{cur}$ and $\mathcal{L}_{next}$ represent the subsets of vertices in the current and next levels of the BFS.
$\mathcal{L}$ stores the BFS level to which each vertex belongs ($-1$ denotes an unvisited vertex). 
The do-while loop (lines 8-16) performs the BFS traversal similar to Algorithm~\ref{algo:dist_RCM}.
The overloaded addition operation of the \Call{SpMSpV}{} at line 12 is set to \emph{min} only for convenience. It can be replaced by any equivalent operation because the order of vertices within a level of BFS does not matter in the discovery of a pseudo-peripheral vertex.
At the end of the BFS, we find a vertex with the minimum degree from the last non-empty level and make it the root for the next BFS (line 17).
The while loop at line 4 terminates when the number of levels in the latest BFS is not greater than the previous BFS.
At this point, Algorithm~\ref{algo:dist_pseudo} returns the pseudo-peripheral vertex $r$.

\section{Distributed memory algorithm}
In order to parallelize the RCM algorithm, we simply need to parallelize the matrix-algebraic functions described in Table~\ref{tab:operations}.

\subsection{Data distribution and storage}
We use the Combinatorial BLAS (CombBLAS) framework~\cite{bulucc2011combinatorial}, which distributes its sparse matrices on a 2D $p_r \times p_c$ processor grid.
Processor $P(i,j)$ stores the submatrix $\mA_{ij}$ of dimensions $(m/p_r)\times (n/p_c)$ in its local memory. 
Vectors are also distributed on the same 2D processor grid. 
CombBLAS supports different formats to store its local submatrices. In this work, we use the CSC format as we found it to be the fastest 
for the SpMSpV operation with very sparse vectors, which is often the case for matrices where RCM is commonly used. 
CombBLAS uses a vector of \{index, value\} pairs for storing sparse vectors. 
To balance load across processors, we randomly permute the input matrix $\mA$ before running the RCM algorithm.



\subsection{Analysis of the distributed algorithm}

We measure communication by the number of {\em words} moved ($W$) and
the number of {\em messages} sent ($S$). The cost of communicating a length $m$ message is $\alpha + \beta m$ where $\alpha$ is the
latency and $\beta$ is the inverse bandwidth, both defined relative to the cost of a single arithmetic operation. Hence, an algorithm that
performs $F$ arithmetic operations, sends $S$ messages, and moves $W$ words takes $T= F + \alpha S + \beta W$ time. 

Since the amount of work is variable across BFS iterations, we analyze the aggregate cost over the whole BFS.
For ease of analysis, matrix and vector nonzeros are assumed to be i.i.d.\ distributed. 
We also assume a square processor grid $p_c{=}p_r{=}\sqrt{p}$. 
Number of complete breadth-first searches is denoted by $\left\vert{\mathrm{iters}} \right\vert$. The value of $\left\vert{\mathrm{iters}} \right\vert$ is exactly one in Algorithm~\ref{algo:dist_RCM}, but it can be more than one in Algorithm~\ref{algo:dist_pseudo}.

We leverage the 2D SpMSpV algorithms implemented in CombBLAS.
The complexity of parallel SpMSpV algorithm has been analyzed in this context recently~\cite{matchingipdps16}, hence we just state
the result here: 
 $$ T_{ \Call{SpMSpV}{}}  = O \Bigl( \frac{m}{p} +  \beta  \bigl ( \frac{m}{p} + \frac{n}{\sqrt{p}} \bigr ) +  \left\vert{\mathrm{iters}} \right\vert \alpha \sqrt{p} \Bigr) .$$

As described before, the \Call{SortPerm}{} function returns the sorted permutation of the vertices in the next frontier $\mathcal{L}_{next}$ based on the lexicographic order of (parent\_label, degree, vertex\_id), where parent\_label is the label of the parent of a vertex in $\mathcal{L}_{next}$.
We observe that parents of $\mathcal{L}_{next}$ are a subset of the current frontier $\mathcal{L}_{cur}$ and vertices in $\mathcal{L}_{cur}$ are incrementally labeled in the previous iteration.
Hence, the number of unique parent labels is less than or equal to $\dnnz(\mathcal{L}_{cur})$. 
This observation motivates us to design a distributed bucket sort algorithm where  the $i$-th processor is responsible for sorting vertices whose parent's labels fall in the range $$[\frac{\dnnz(\mathcal{L}_{cur})}{p}i, \frac{\dnnz(\mathcal{L}_{cur})}{p} (i+1)).$$ 
Hence, processors exchange tuples using AllToAll and the responsible processors sort tuples locally. 
After sorting, the vertex indices associated with the tuples work as the inverse permutation. Processors conduct another round of AllToAll (only the indices) to obtain the sorted permutation. 

The total number
of nonzeros sorted is exactly the sum of the frontier sizes over all iterations, which is $O(n)$.  
Hence the per-process cost of \Call{SortPerm}{} is upper bounded by
$$ T_{\Call{SortPerm}{}}= O \Bigl( \frac{n \log{n}}{p} +  \beta  \frac{n}{p} +  \left\vert{\mathrm{iters}} \right\vert \alpha p \Bigr) ,$$
 using personalized all-to-all~\cite{bruck1997efficient}. We found our specialized bucket sort to be faster than state-of-the-art 
  general sorting libraries, such as HykSort~\cite{sundar2013hyksort}.


\section{Results}
\label{sec:results}

\begin{figure}[t!]
\begin{center}
\scalebox{1}{
\footnotesize
\renewcommand{\arraystretch}{1.1}
\addtolength{\tabcolsep}{-5pt}
\begin{tabular}{|lccc|}
	\hline 
\textbf{Name}		& \multirow{3}{*}{Spy Plot} 			& Dimensions		& BW (pre-RCM) 		\\ 
Description		&							& Nonzeros		& BW (post-RCM)	\\
			&							& 		& Pseudo-diameter		\\
	\hline
	\hline
    			& \multirow{4}{*}{\includegraphics[scale=0.024]{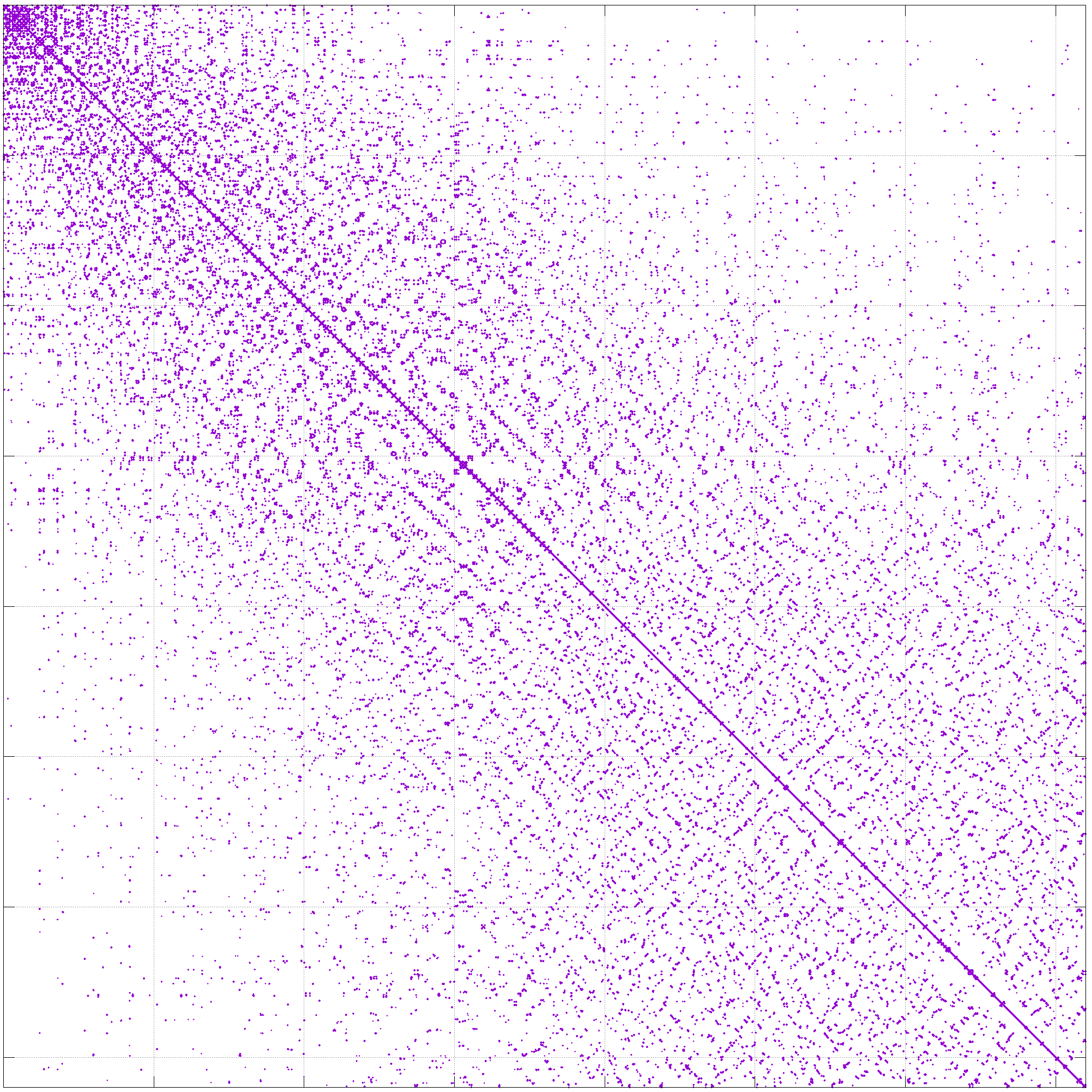}}	& 			&			\\
\textbf{nd24k}	& 								& 72K$\times$72K & 	68,114 		\\
3D mesh  &							& 29M		& 	10,294	\\
problem						& 		&	& 14		\\
	\hline

			& \multirow{4}{*}{\includegraphics[scale=0.12]{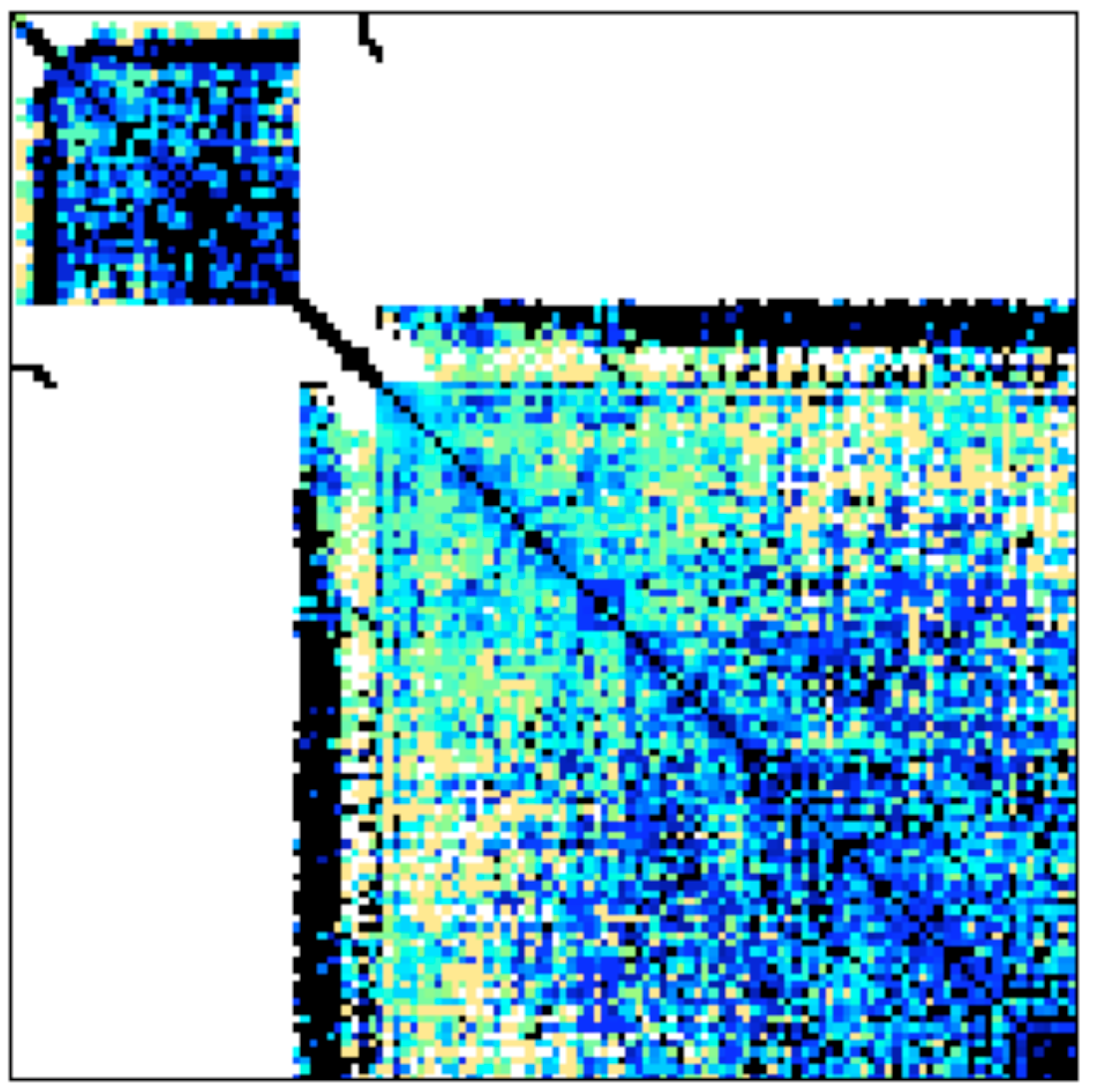}}	& 			&			\\
\textbf{Ldoor}	& 								& 952K$\times$952K	& 	686,979 		\\
structural prob.	&							& 42.49M		& 	9,259	\\
			&							& 		&	178		\\
	
	\hline
	
			& \multirow{4}{*}{\includegraphics[scale=0.10]{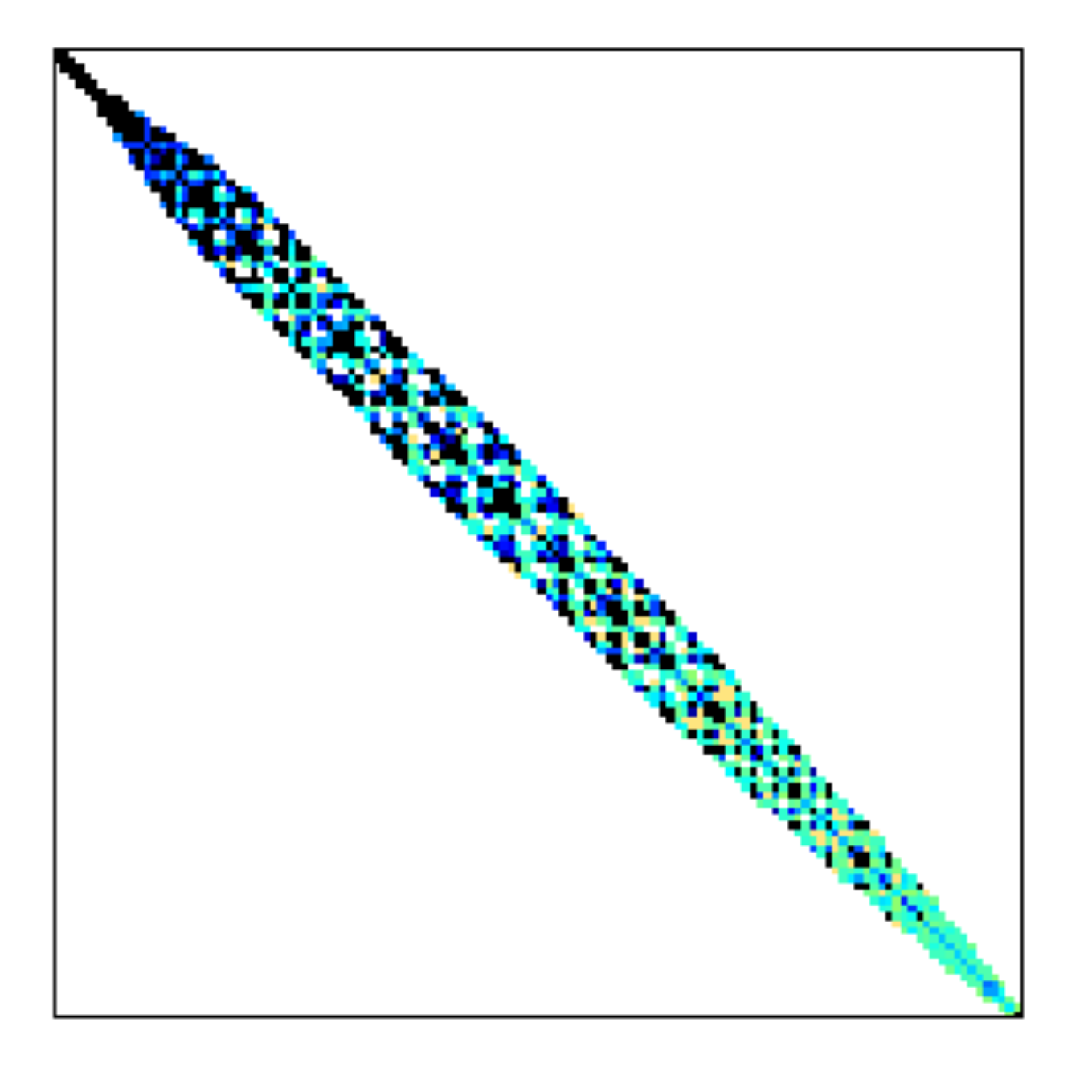}}	&			&			\\
\textbf{Serena}	& 							& 1.39M$\times$1.39M	& 	81,578		\\
gas reservoir  	&								& 64.1M		& 	81,218	\\
simulation		&								& 		&	58		\\
	\hline
	
	\hline

			& \multirow{4}{*}{\includegraphics[scale=0.024]{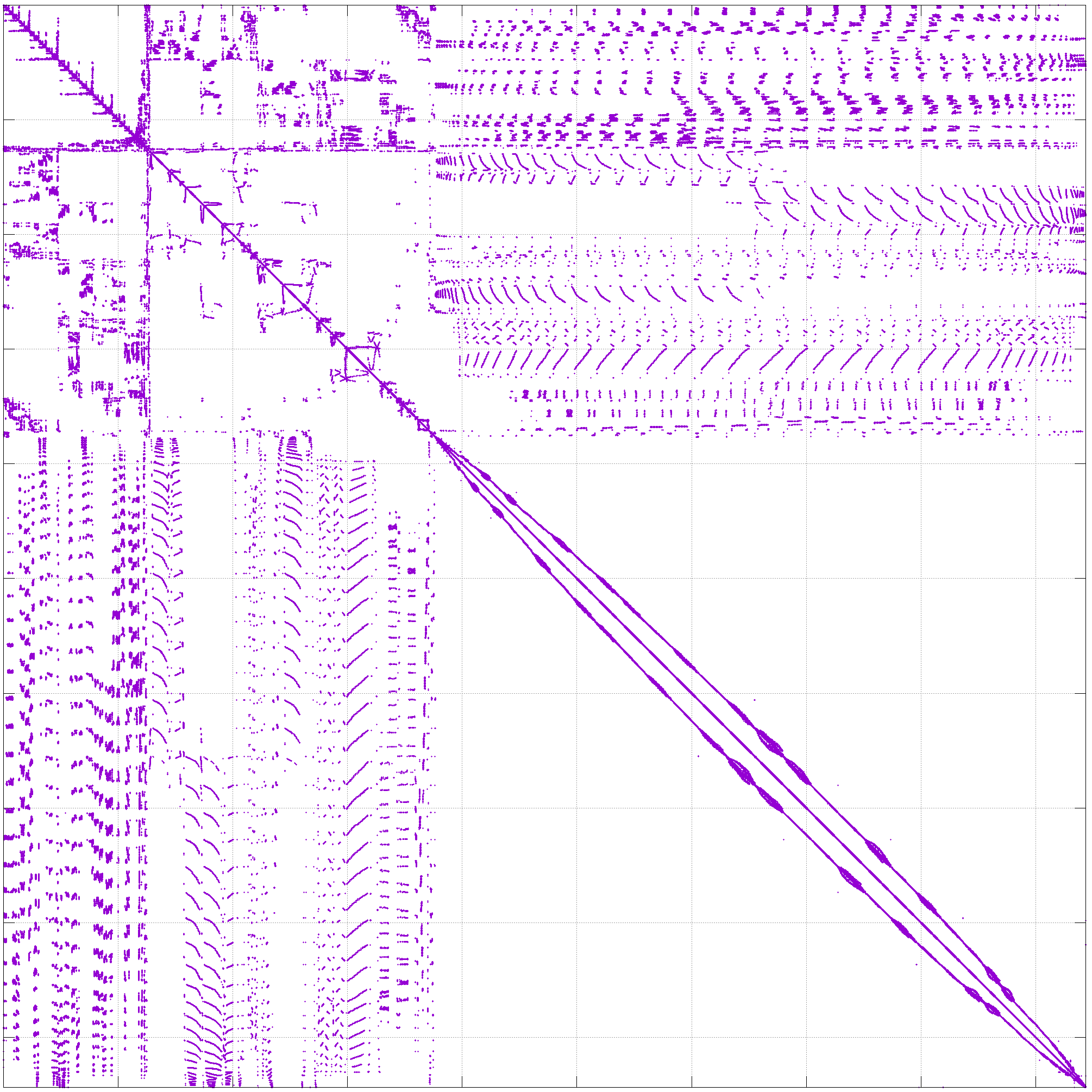}}	& 			&			\\
\textbf{audikw\_1}	& 								& 943K$\times$943K & 	925,946 		\\
 structural prob &							& 78M		& 	35,170	\\
						& 		&	& 82		\\
 \hline
 
 			& \multirow{4}{*}{\includegraphics[scale=0.10]{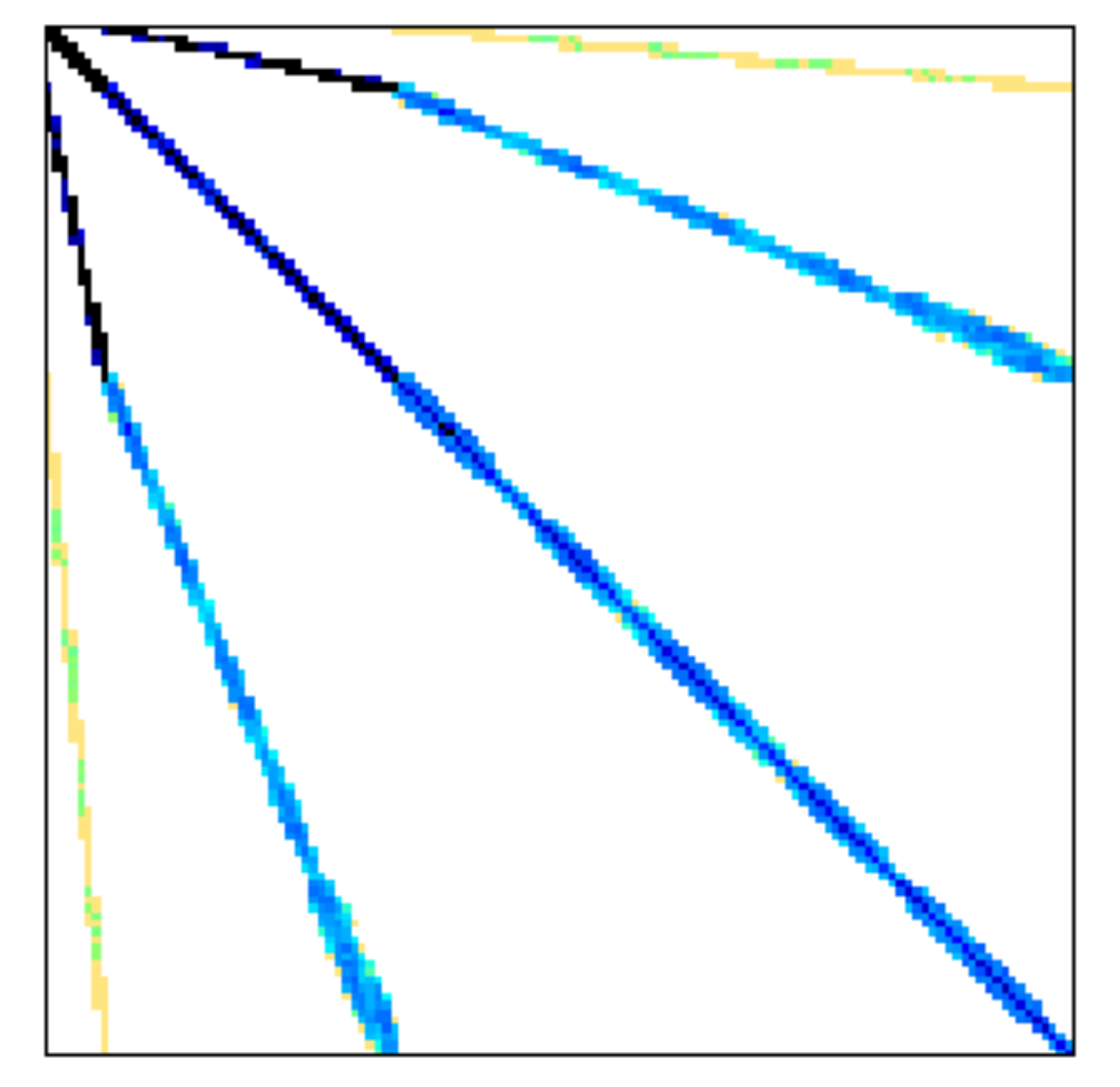}}	&			&			\\
\textbf{dielFilterV3real}	& 							& 1.1M$\times$1.1M	& 	1,036,475		\\
higher-order  	&								& 89.3M		& 	23,813	\\
finite element		&								& 		&	84		\\
\hline

		& \multirow{4}{*}{\includegraphics[scale=0.10]{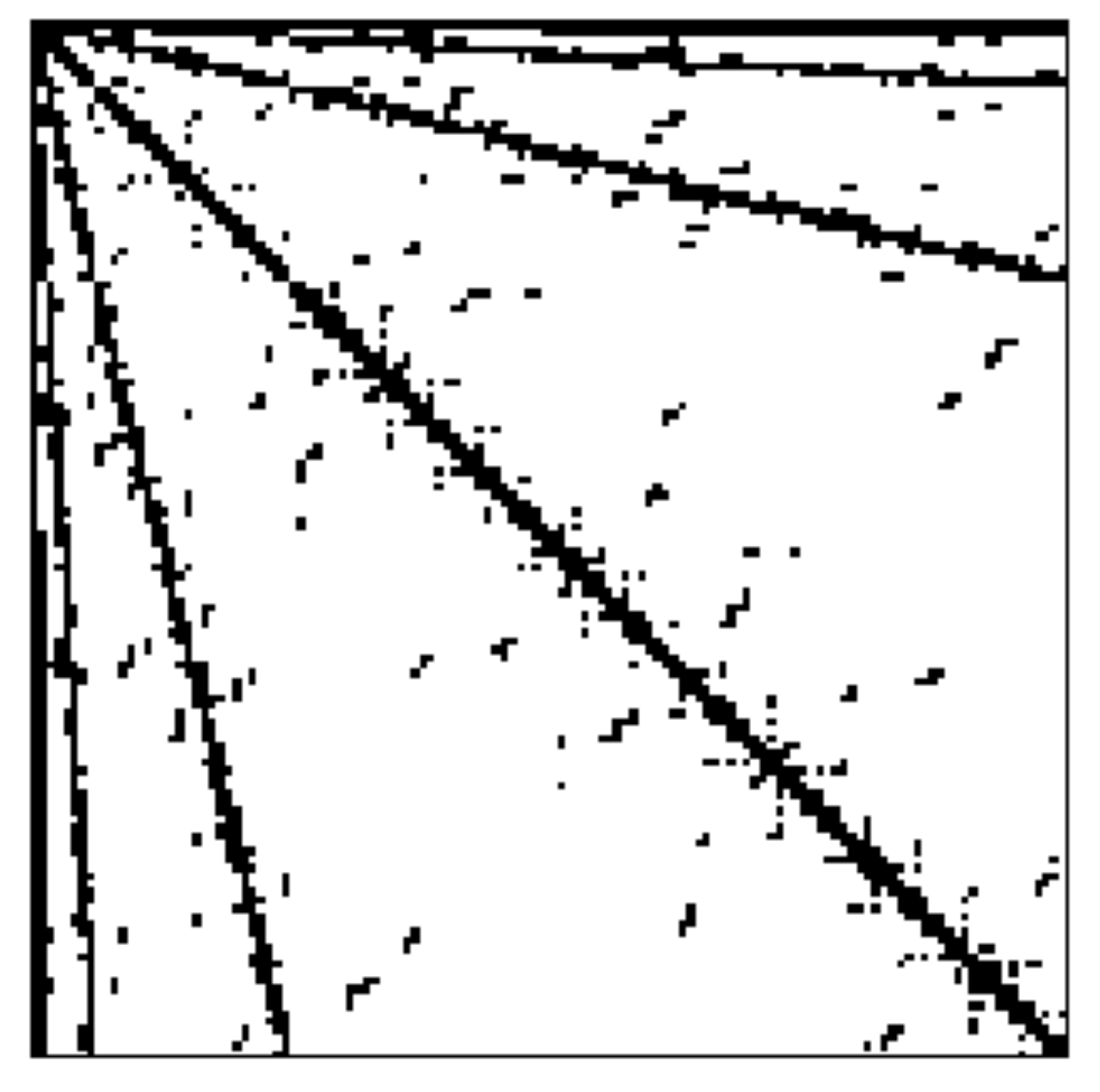}}	&			&			\\

\textbf{Flan\_1565}	& 							& 1.6M$\times$1.6M	& 	20,702		\\
3D model of   	&								& 114M		& 	20,600 	\\
a steel flange		&								& 		&	199		\\
	\hline

		& \multirow{4}{*}{\includegraphics[scale=0.024]{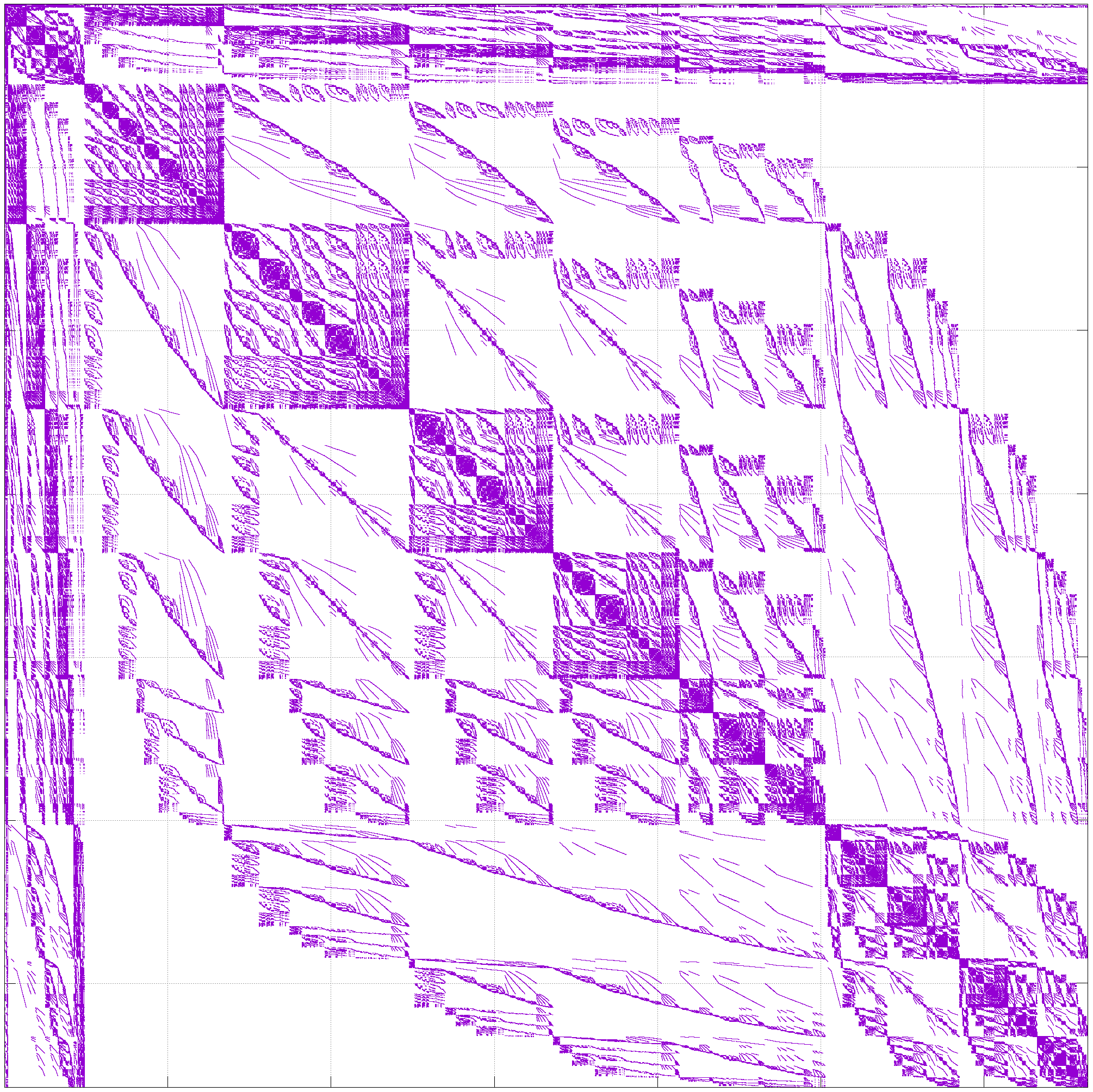}}	&			&			\\
\textbf{Li7Nmax6}	& 							& 664K$\times$664K	& 	663,498		\\
nuclear configuration  	&								& 212M		& 	490,000	\\
interaction calculations		&								& 		&		7	\\
	\hline
    
		& \multirow{4}{*}{\includegraphics[scale=0.024]{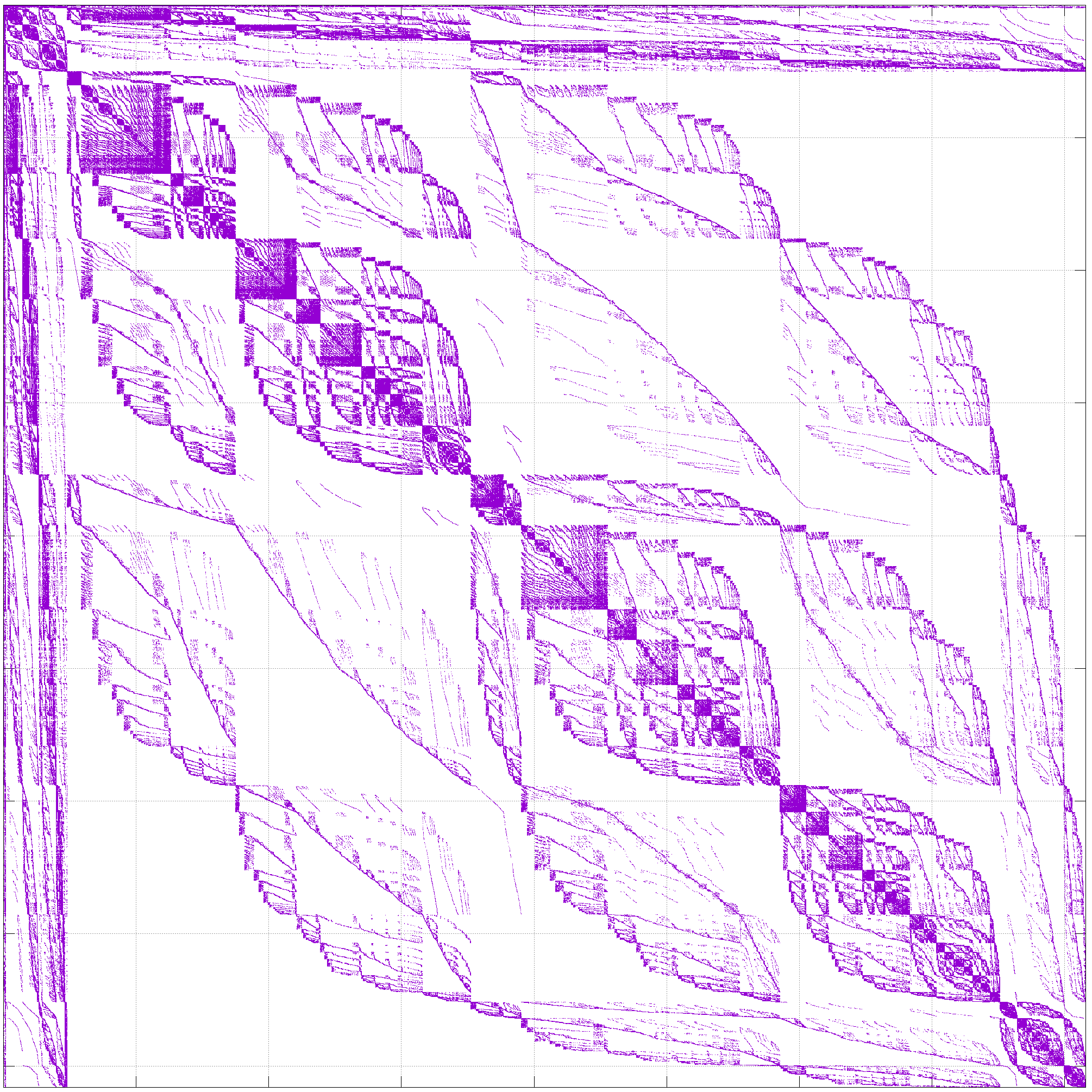}}	&			&			\\
\textbf{Nm7}	& 							& 4M$\times$4M	&  4,073,382	\\
nuclear configuration  	&								& 437M		& 	3,692,599	\\
interaction calculations		&								& 		&		5	\\
	\hline
	
		& \multirow{4}{*}{\includegraphics[scale=0.10]{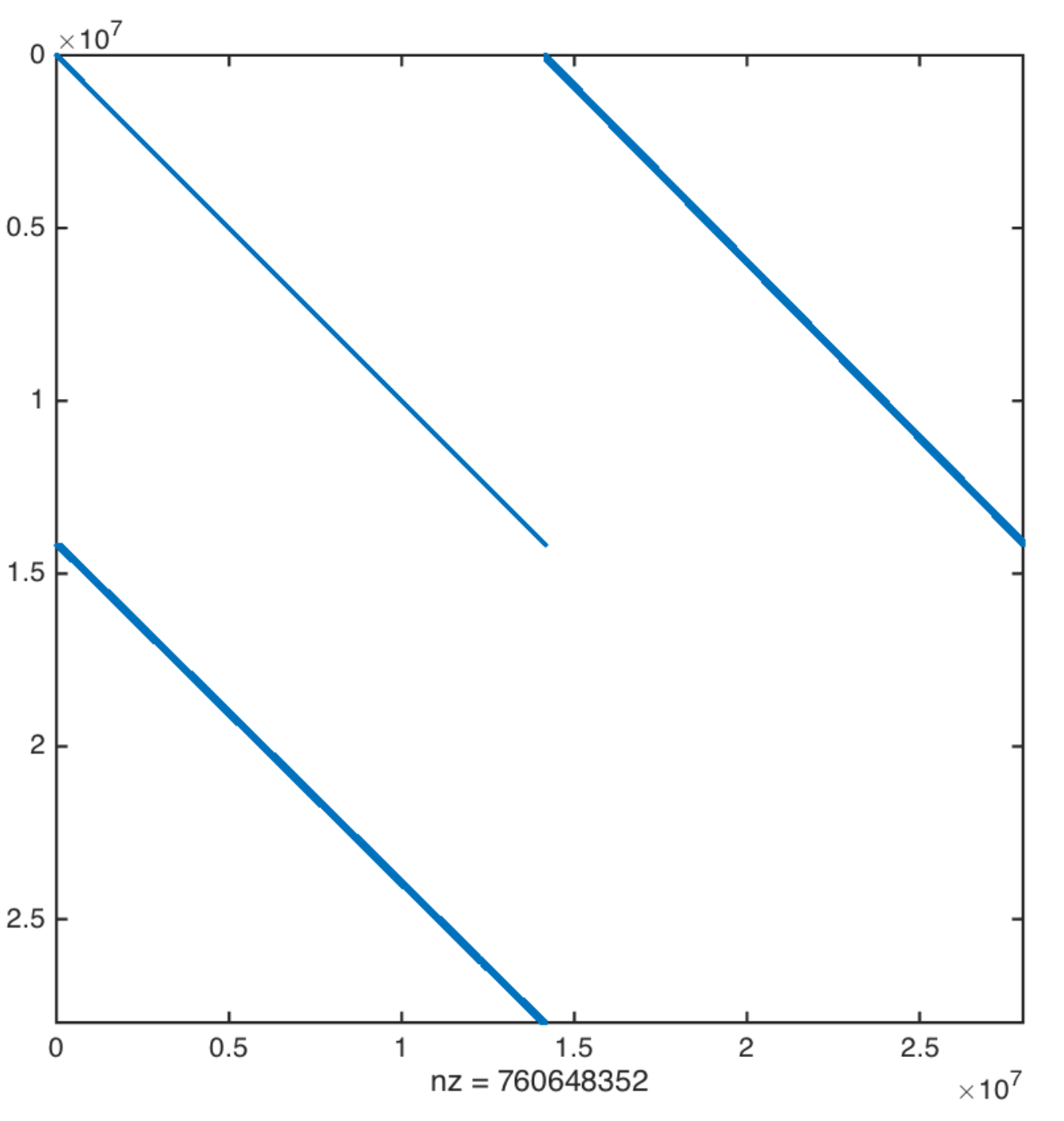}}	&			&			\\
\textbf{nlpkkt240}	& 							& 78M$\times$78M	& 	14,169,841		\\
Sym.\ indefinite   	&								& 760M		& 	361,755	\\
KKT matrix 		&								& 		&	243		\\
	\hline

\end{tabular}}
\end{center}
\caption{\small \bf Structural information on the sparse matrices used in our
  experiments. All matrices, except two,
  are from the University of Florida sparse
  matrix collection~\cite{ufget}. Li7Nmax6 and Nm7~\cite{ipdps14} are from nuclear configuration interaction calculations. \label{fig:testsuite}
  }
\end{figure}

\subsection{Experimental Platform}
We evaluated the performance of parallel RCM algorithm on Edison, a Cray XC30 supercomputer at NERSC.
In Edison, nodes are interconnected with the Cray Aries network using a Dragonfly topology. 
Each compute node is equipped with 64 GB RAM and two 12-core 2.4 GHz Intel Ivy Bridge processors, each with 30 MB L3 cache.
We used Cray's MPI implementation, which is based on MPICH2. 
We used OpenMP for intra-node multithreading and compiled the code with gcc~5.2.0 with \texttt{\mbox{\small -O2 -fopenmp}} flags. 
In our experiments, we only used square process grids because rectangular grids are not supported in CombBLAS~\cite{bulucc2011combinatorial}. 
When $p$ cores were allocated for an experiment, we created a $\sqrt{p/t} \times \sqrt{p/t} $ process grid, where $t$ was the number of threads per process.
In our hybrid OpenMP-MPI implementation, all MPI processes performed local computation followed by synchronized communication rounds. 
Local computation in every matrix-algebraic kernel was fully multithreaded using OpenMP.
Only one thread in every process made MPI calls in the communication rounds. 

\begin{table}[!t]
 \centering

 \caption{ The bandwidth and runtime of the shared-memory RCM implementation in SpMP. SpMP did not finish in 30 minutes for \texttt{Nm7}. On \texttt{Nm7} and \texttt{nlpkkt240} matrices, our distributed-memory implementation ran out of memory on a single node of Edison. 
\label{table:SpMP}
}
 \scalebox{0.9}{
 \begin{tabular}{@{} l r r  r   r r  r   r  @{}}
    \toprule
 Graphs          & \multicolumn{4}{c}{SpMP} & \multicolumn{3}{c}{Distributed RCM}\\
	                & BW      & \multicolumn{3}{c}{Runtime (sec)} & \multicolumn{3}{c}{Runtime (sec)}\\
 	                 &         & 1t    & 6t    & 24t    & 1t    & 6t   & 24t\\
  \midrule
nd24k              & 10,608  & 0.26  & 0.06  & 0.03   & 1.45  & 0.38 & 0.12 \\
ldoor              & 9,099   & 3.25  & 0.52  & 0.28   & 4.63  & 1.52 & 0.74 \\
Serena             & 85,229  & 1.64  & 0.49  & 0.66   & 7.75  & 2.26 & 1.08\\
audikw\_1          & 34,202  & 1.31  & 0.34  & 0.16   & 7.31  & 1.99 & 0.81\\
dielFilterV3real   & 25,436  & 1.99  & 0.73  & 0.46   & 8.63  & 2.37 & 0.95\\
Flan\_1565         & 20,849  & 1.86  & 0.44  & 0.17   & 12.11 & 3.88 & 1.35 \\
Li7Nmax6           & 443,991 & 4.62  & 1.48  & 0.87   & 20.28 & 4.91 & 2.85 \\
Nm7                & -       & -     & -     & -      & -     & -    & - \\
nlpkkt240          & 346,556 & 57.21 & 25.17 & 9.92   & -     & -    & - \\
  \toprule
  \end{tabular}}
   \vspace{-6pt}
 \end{table}

\subsection{Matrix Suite}
The sparse matrix test suite used in our experiments are shown in Figure~\ref{fig:testsuite}.
These matrices came from a set of real applications, where either a sparse system $Ax=b$ is solved
or an eigenvalue problem $Ax = \lambda x$ is solved. The matrices were chosen to represent a variety of different structures and nonzero densities.  
Since RCM is only well defined on symmetric matrices, all matrices are symmetric.

In the last column of Figure~\ref{fig:testsuite}, the original (pre-RCM) as well as final (post-RCM) bandwidth of the matrix 
are shown. In the majority of the cases, RCM effectively reduces the bandwidth. 
\texttt{Serena} and \texttt{Flan\_1565} seem to be the only two matrices where RCM was ineffective in that regard.

\subsection{Shared-memory performance}
Our implementation is fully multithreaded to take advantage of the shared-memory parallelism available within a node of modern supercomputers.
Here, we compare the quality and runtime of our algorithm with the RCM
implementation in SpMP (Sparse Matrix Pre-processing) by Park et al.~\cite{spmp}, which is based on optimization from~\cite{spmp-optim} and on the algorithm presented in~\cite{spmp-rcm}.
The results from SpMP is shown in Table~\ref{table:SpMP}.
For four out of eight matrices where SpMP was able to compute RCM, the RCM ordering from our distributed-memory algorithm (shown in Figure~\ref{fig:testsuite}) yields smaller bandwidths than SpMP.
SpMP is faster than our implementation in shared-memory due to our distributed-memory parallelization overheads.
However, SpMP sometimes loses efficiency across NUMA domains. For example, SpMP slows down for \texttt{Serena} on 24 cores compared to 6 cores.
 

\begin{figure*}[!t]
   \centering
   
   \subfloat{\includegraphics[scale=.29]{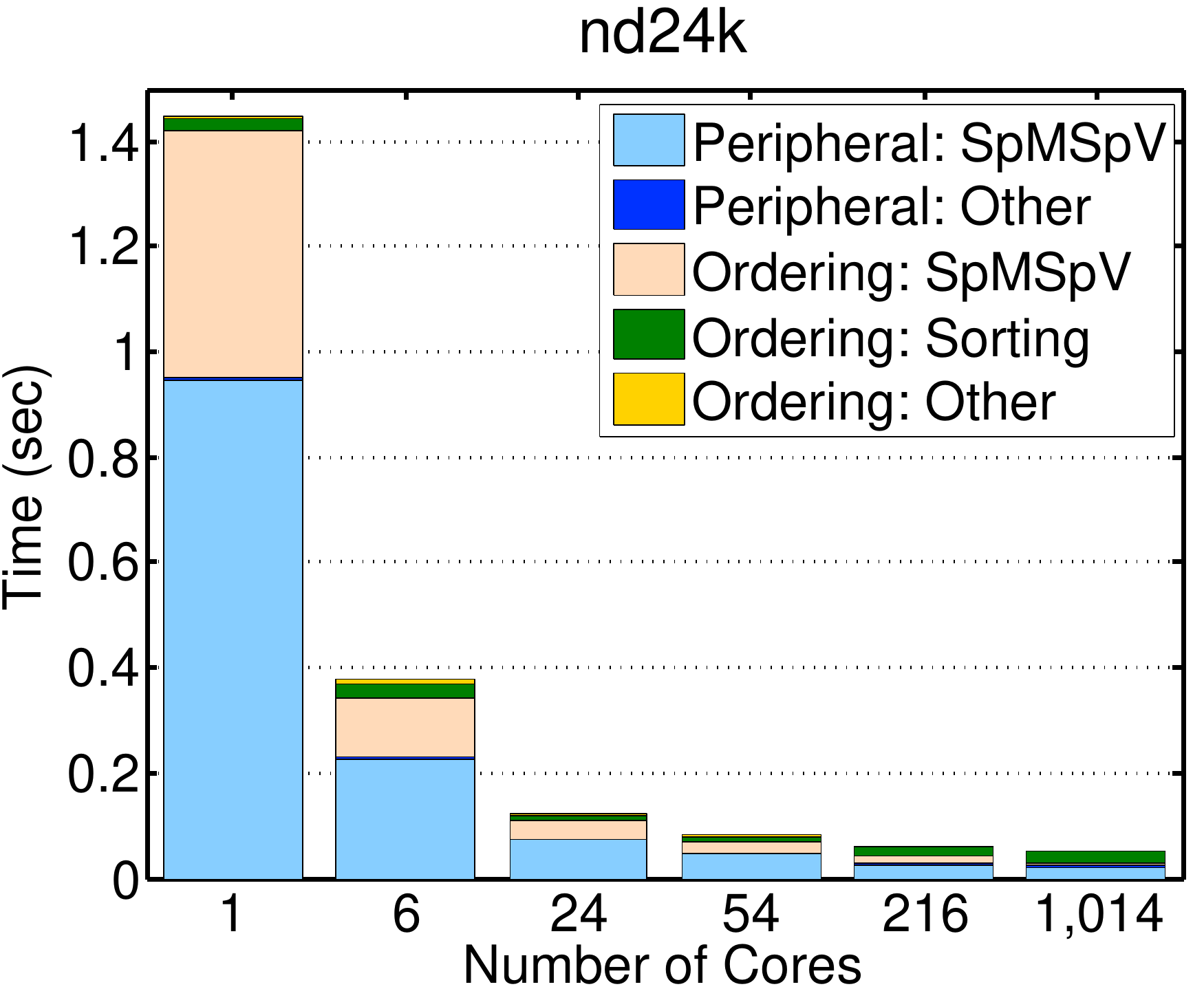}}
   ~ ~
    \subfloat{\includegraphics[scale=.29]{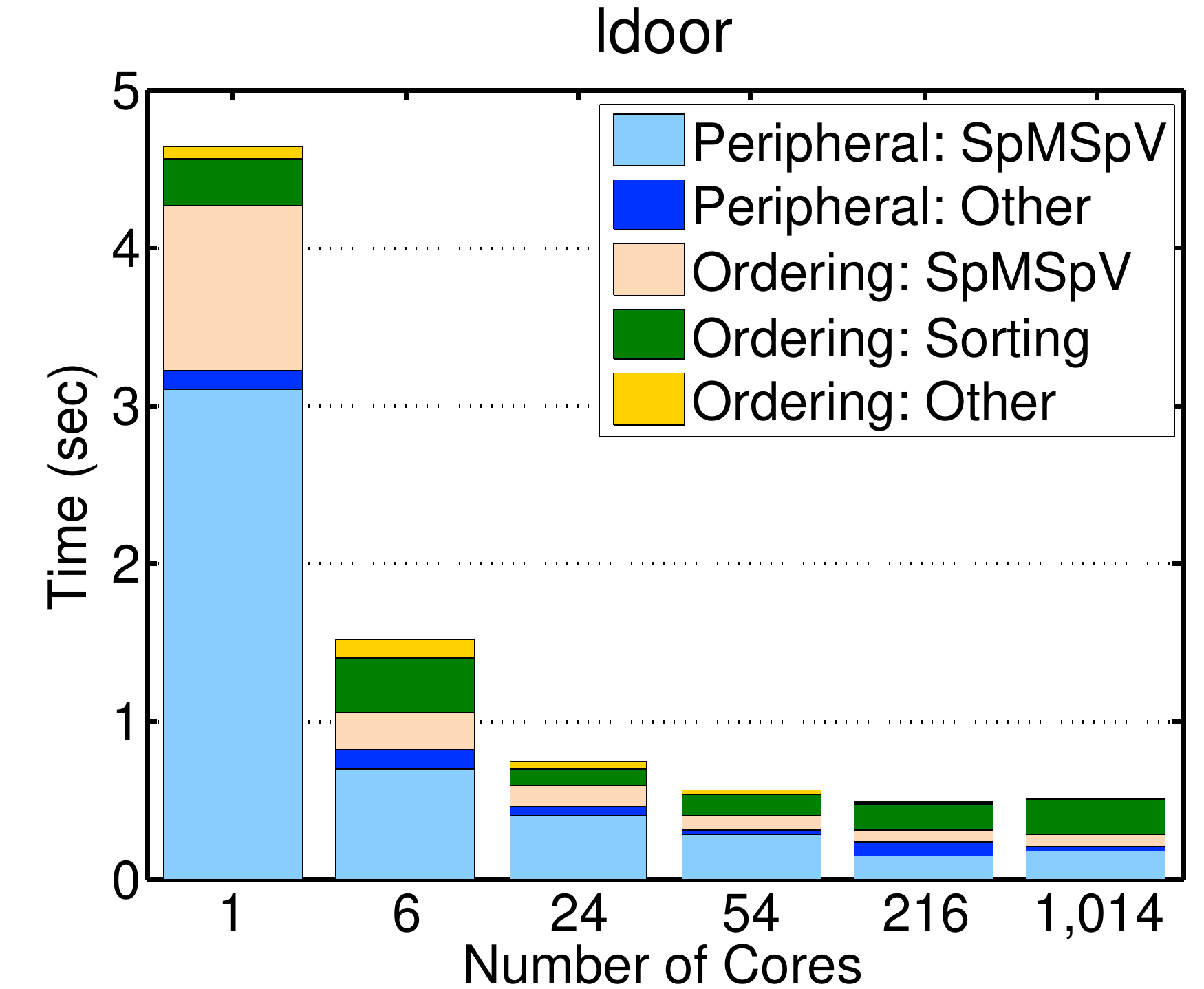}} 
   ~ ~
   \subfloat{\includegraphics[scale=.29]{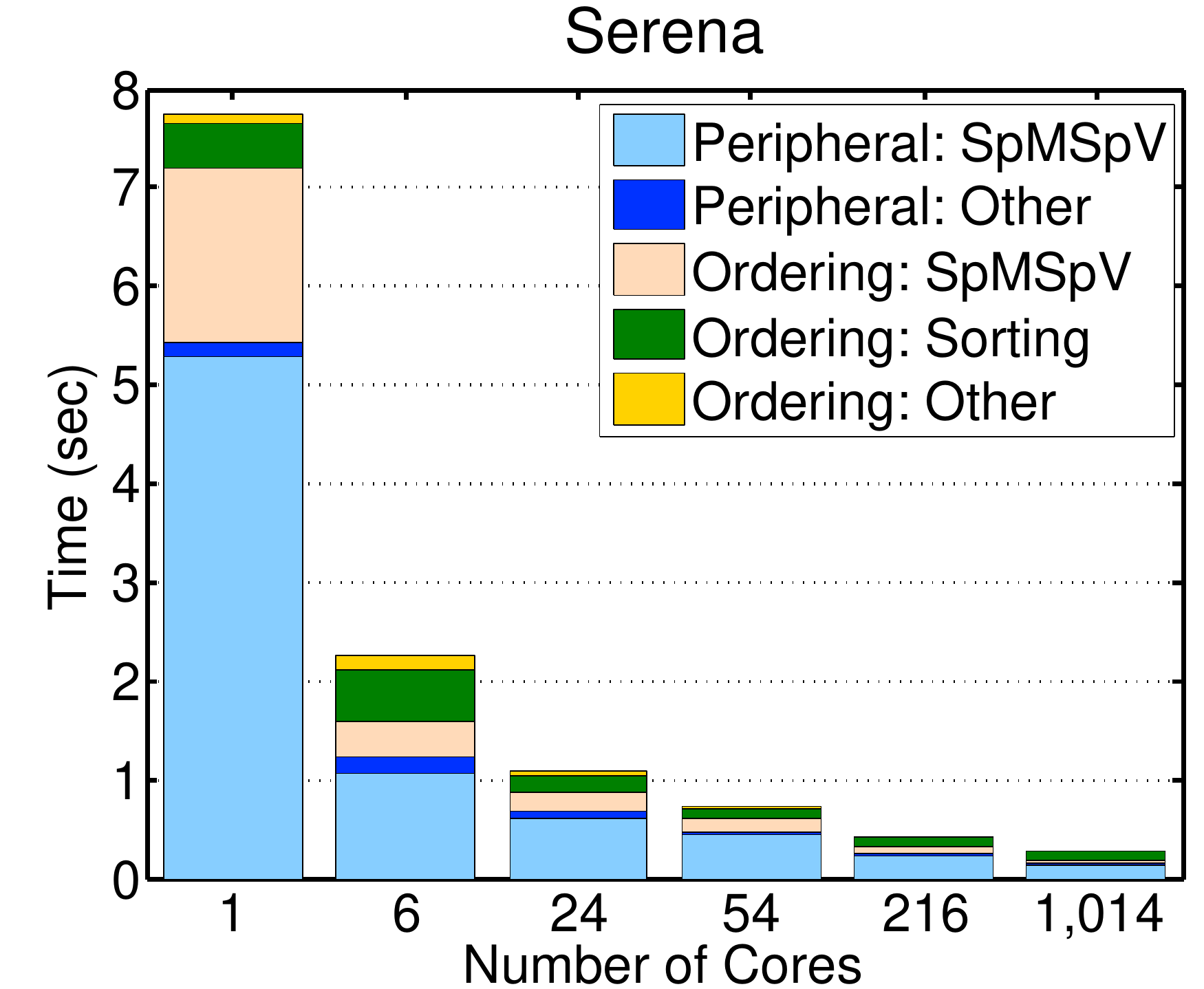}}
   \\
   \subfloat{\includegraphics[scale=.29]{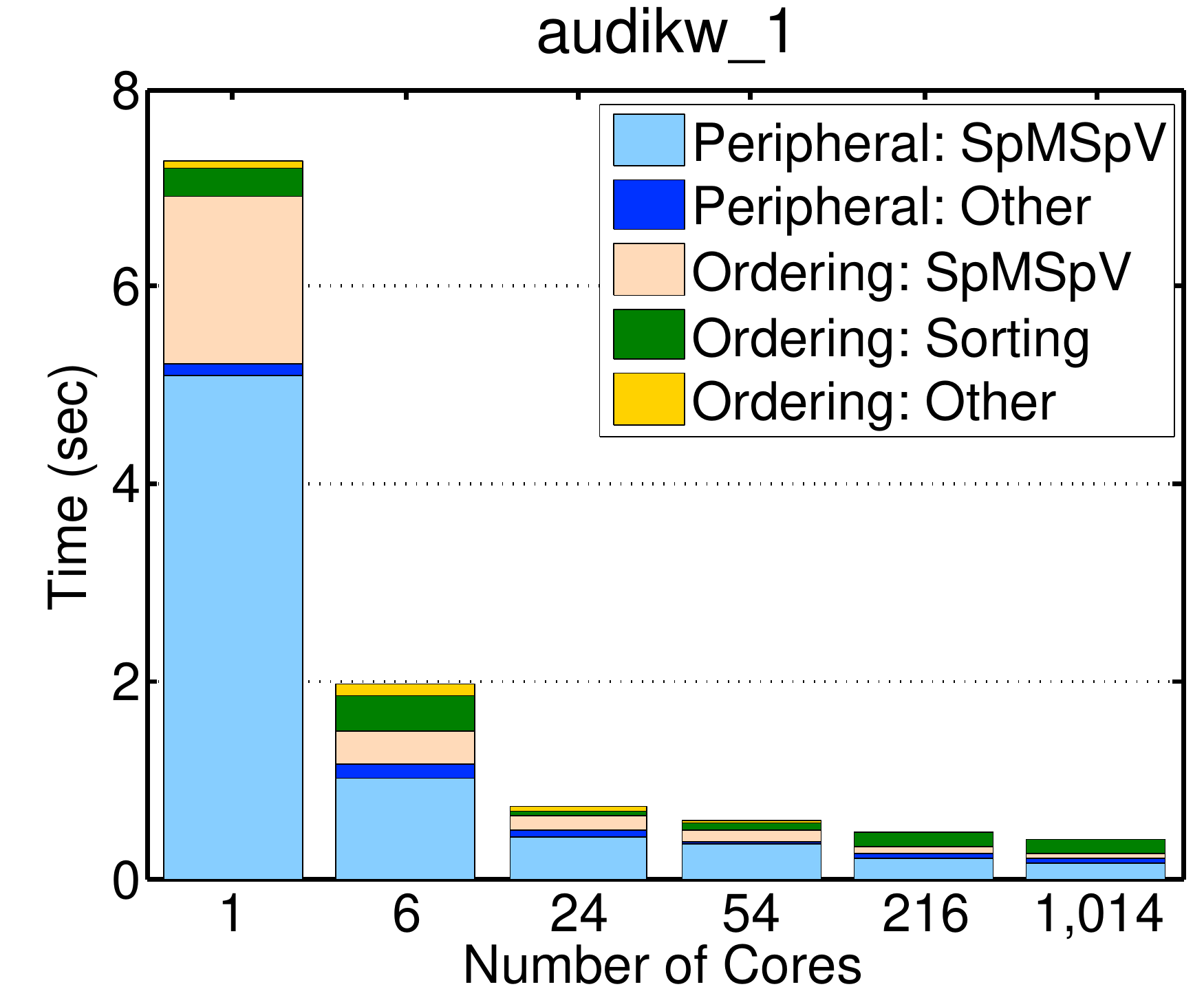}} 
   ~ ~
   \subfloat{\includegraphics[scale=.29]{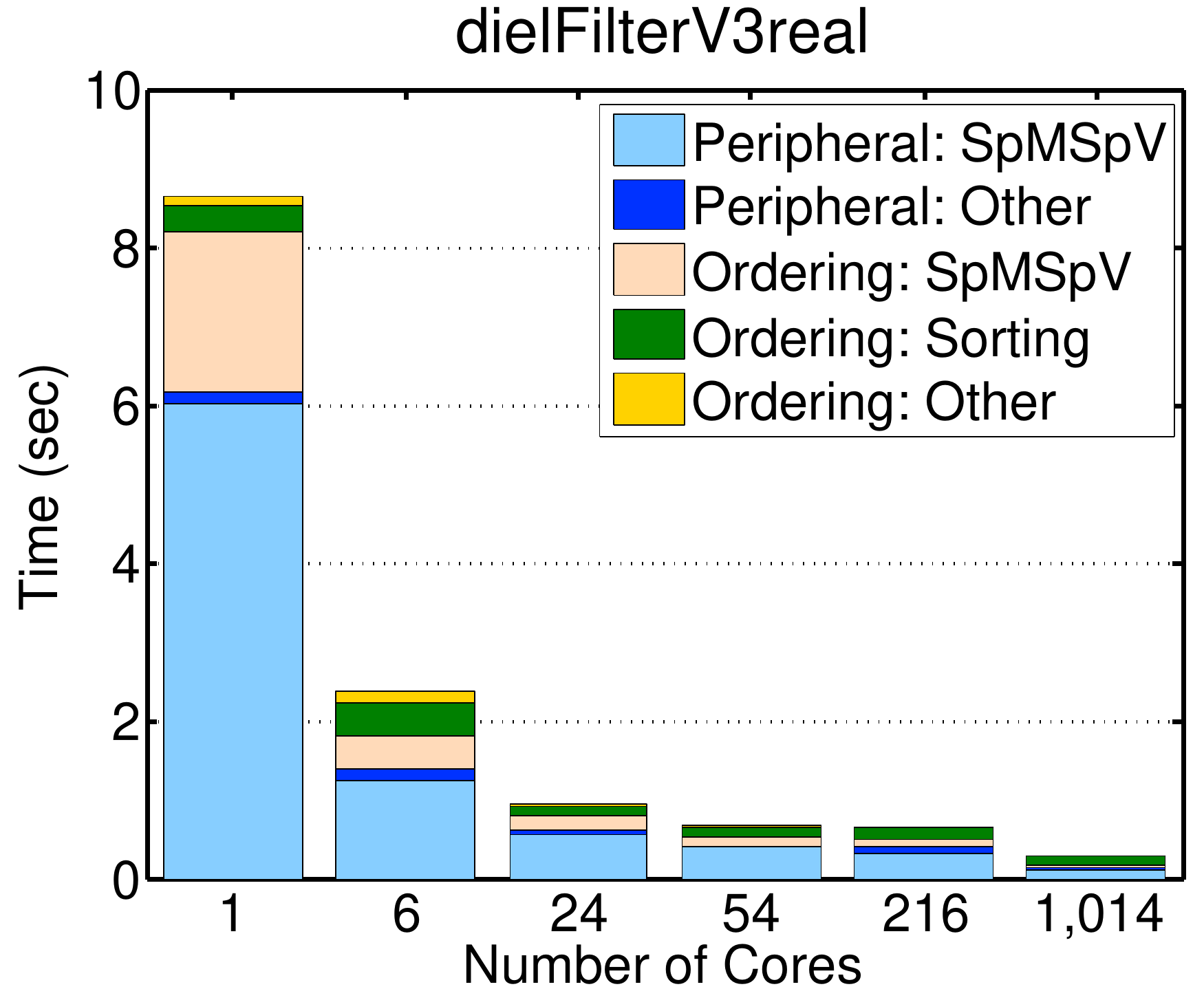}}
   ~ ~
   \subfloat{\includegraphics[scale=.29]{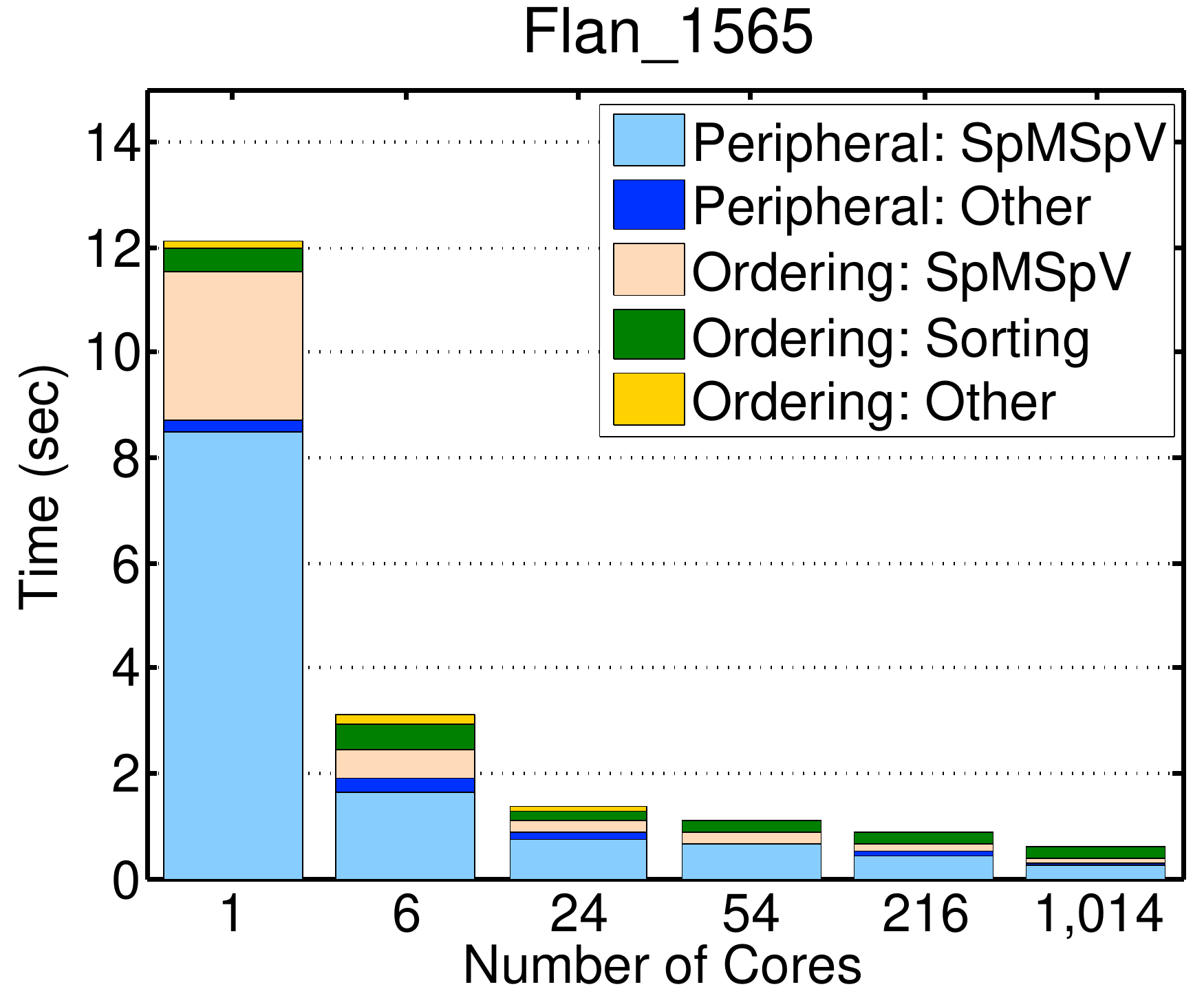}}
  \\

   \subfloat{\includegraphics[scale=.29]{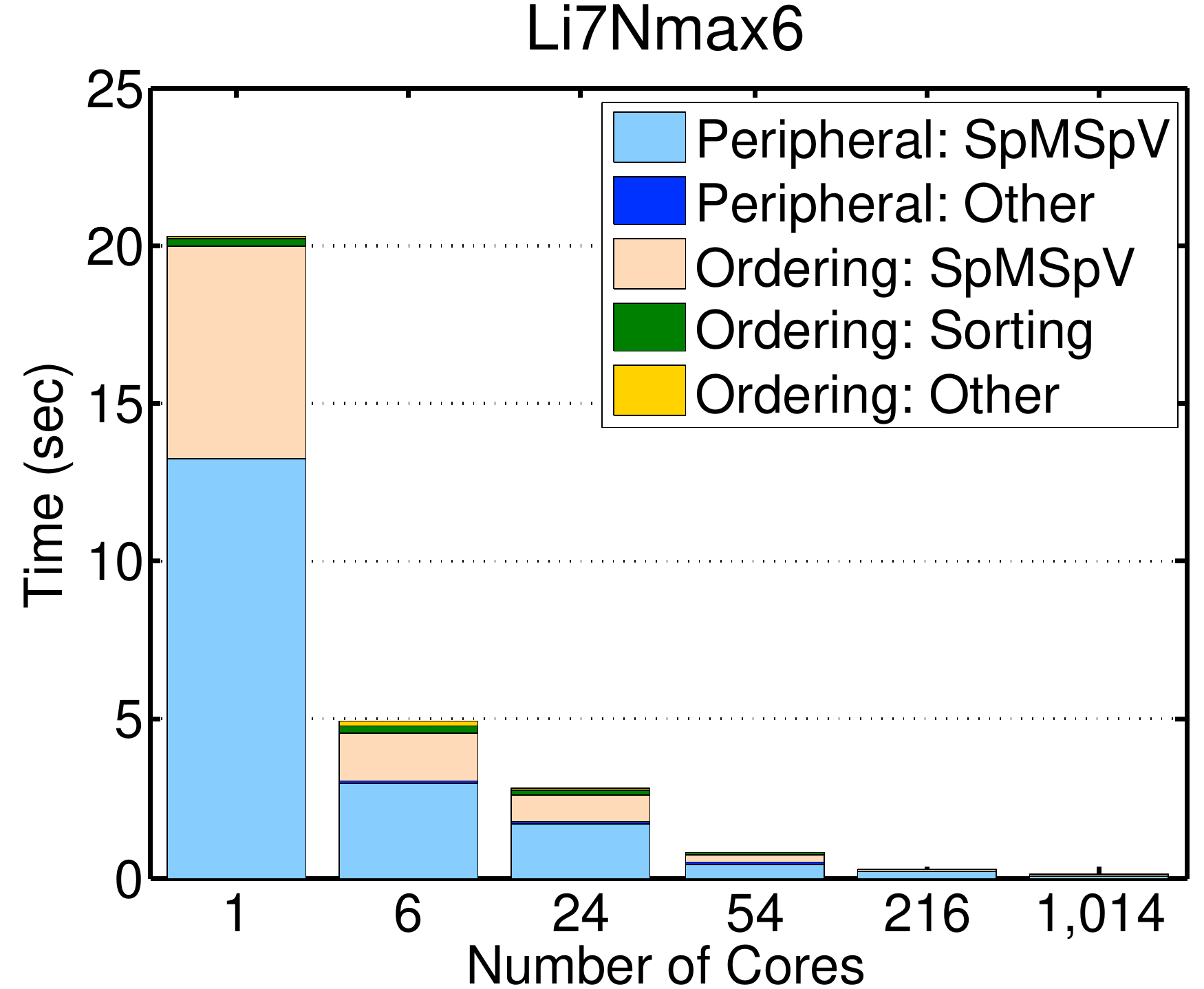}}
   ~~
   \subfloat{\includegraphics[scale=.29]{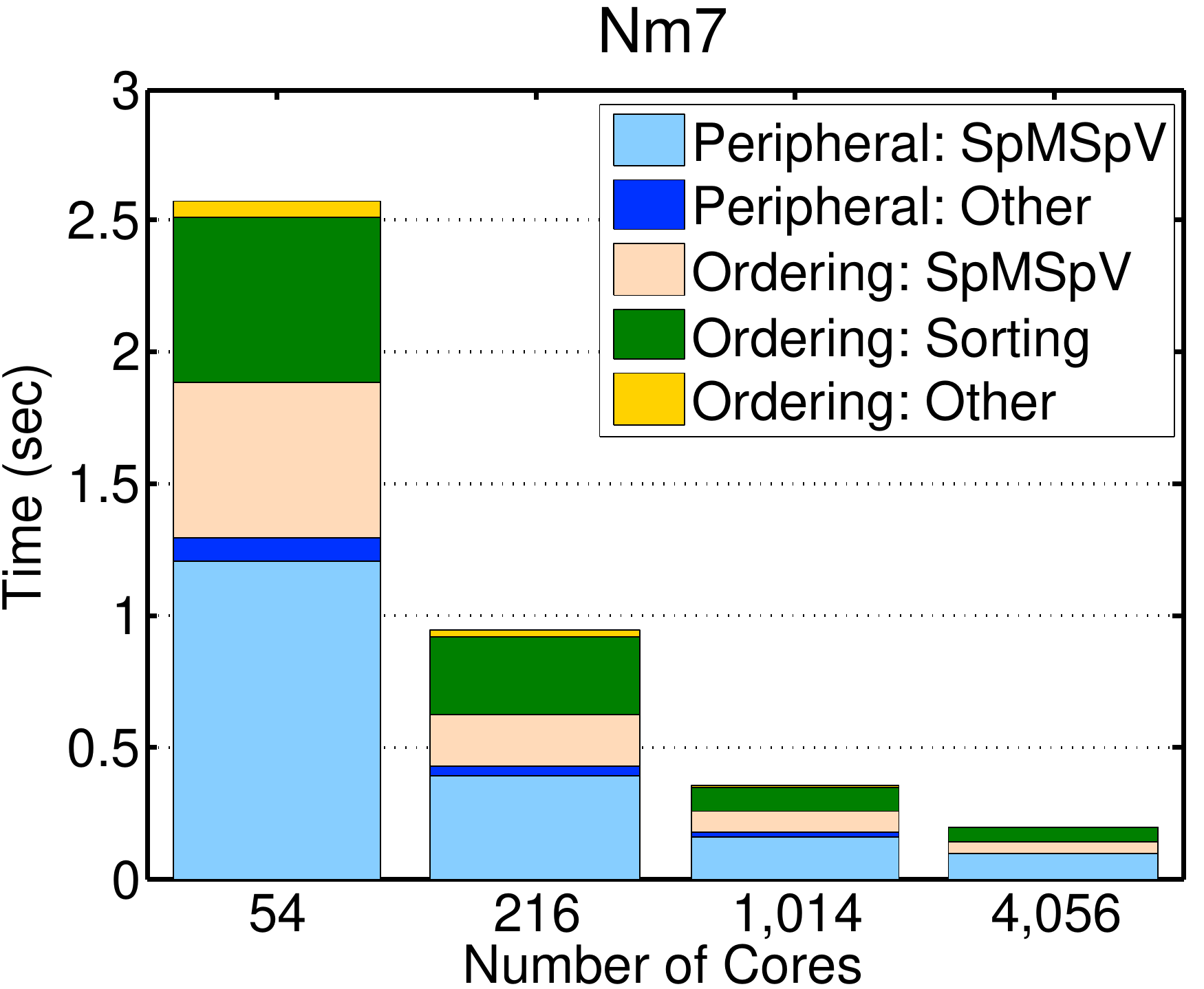}}
   ~ ~
   \subfloat{\includegraphics[scale=.29]{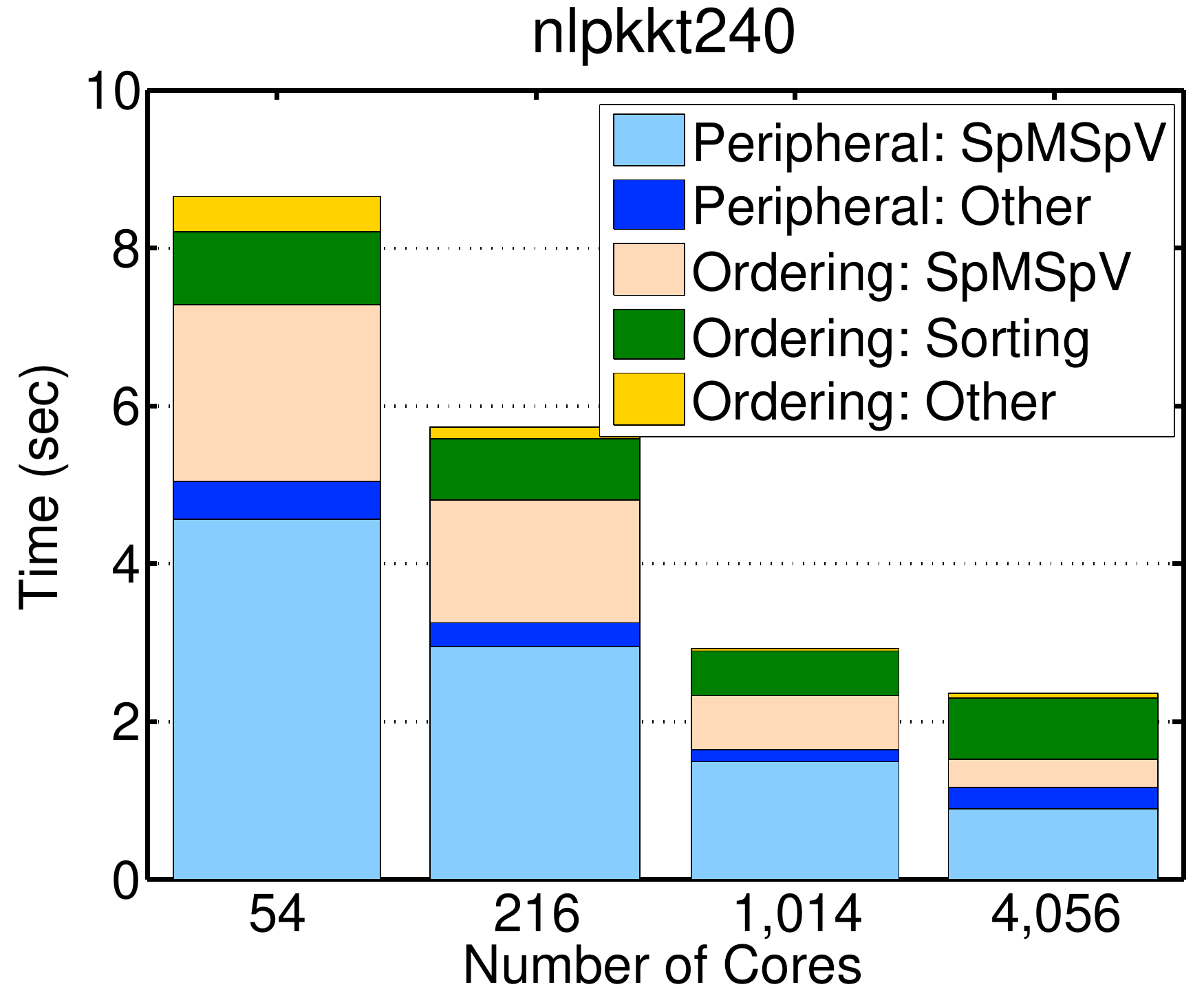}}
   
   \caption{Runtime breakdown of  distributed-memory RCM on different graphs on Edison. Six threads per MPI process are used when the number of cores is greater than or equal to six. \texttt{Nm7} and \texttt{nlpkkt240} ran out of memory on a single node of Edison.}
   \label{fig:timebreak_6t}
\end{figure*}

One very important thing to note is that in order to compute an ordering using a serial or multithreaded implementation of the RCM algorithm such as SpMP, the matrix structure has to be gathered on a single node. Indeed, in many real-life applications, the matrix is already distributed and this mandatory communication step has a non-negligible cost. For example, it takes over 9 seconds to gather the \texttt{nlpkkt240} matrix from being distributed over 1024 cores into a single node/core.
This time is approximately $3\times$ longer than
computing RCM using our algorithm on the same
number of cores.
One of the key benefit of our approach is that it does not require this step.
Adding those times to the time required to compute the ordering itself using SpMP makes our approach highly competitive and often faster.

\begin{figure*}[!t]
   \centering
      
   \subfloat{\includegraphics[scale=.3]{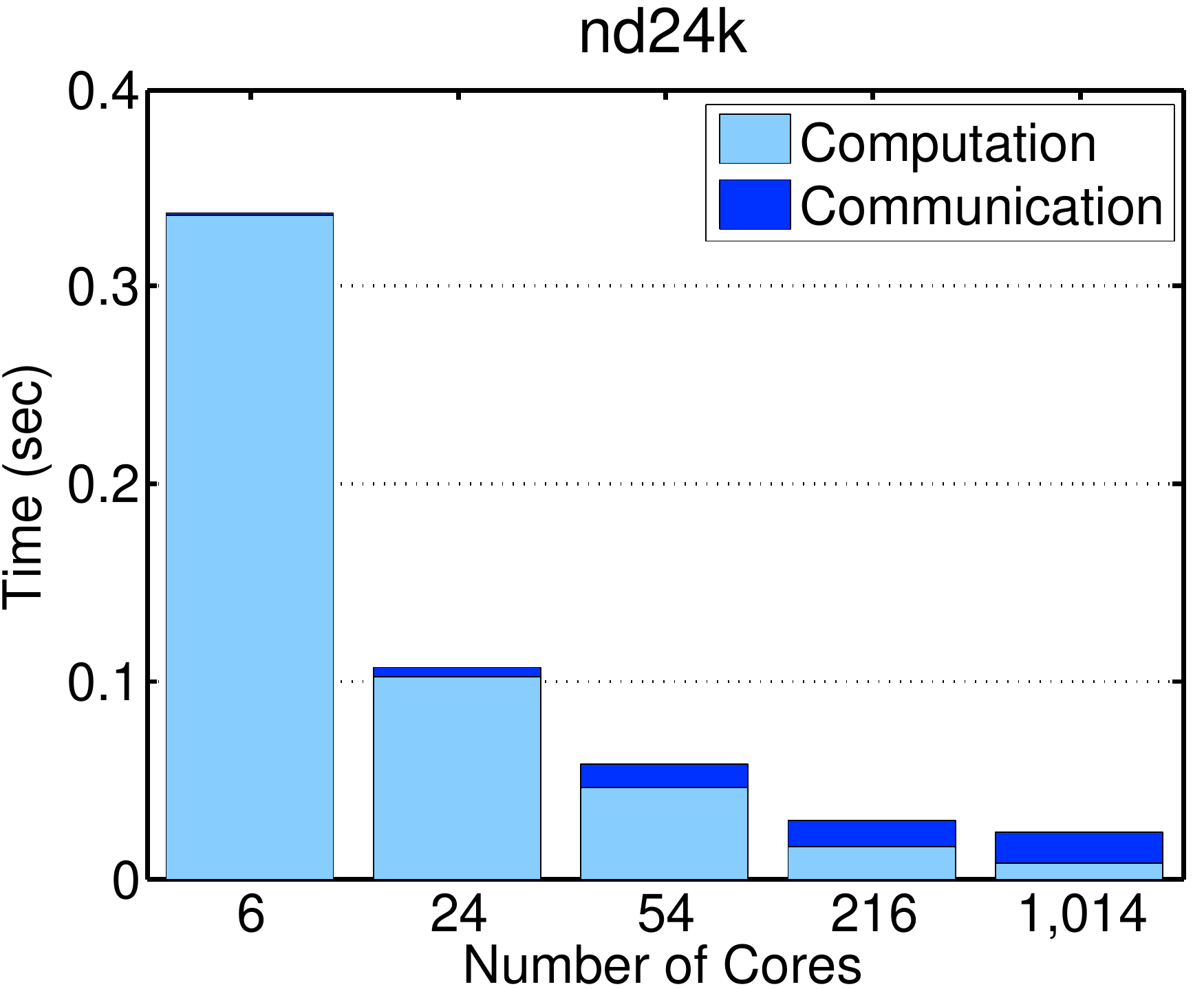}}
   ~ ~
    \subfloat{\includegraphics[scale=.3]{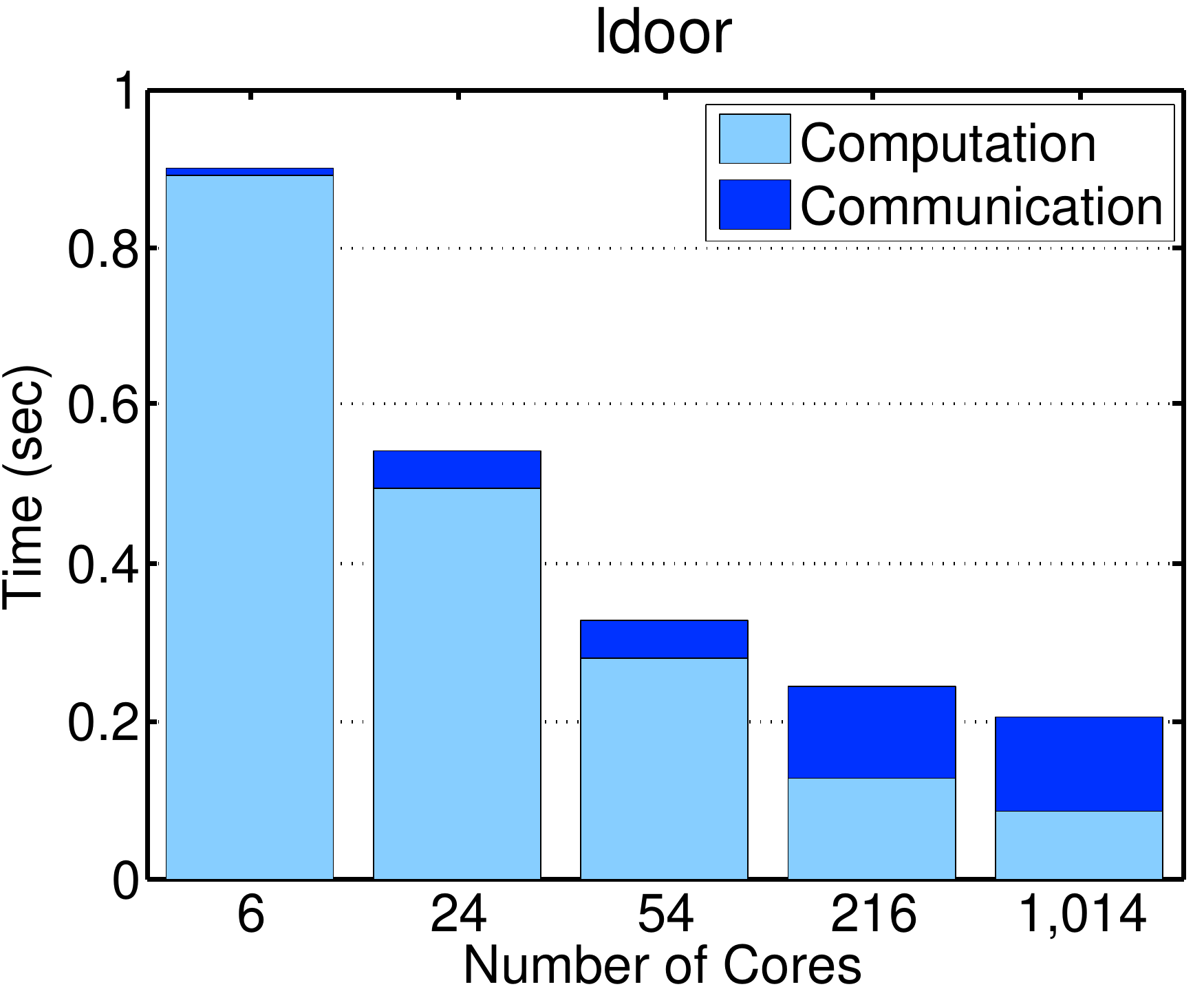}} 
   ~ ~
   \subfloat{\includegraphics[scale=.3]{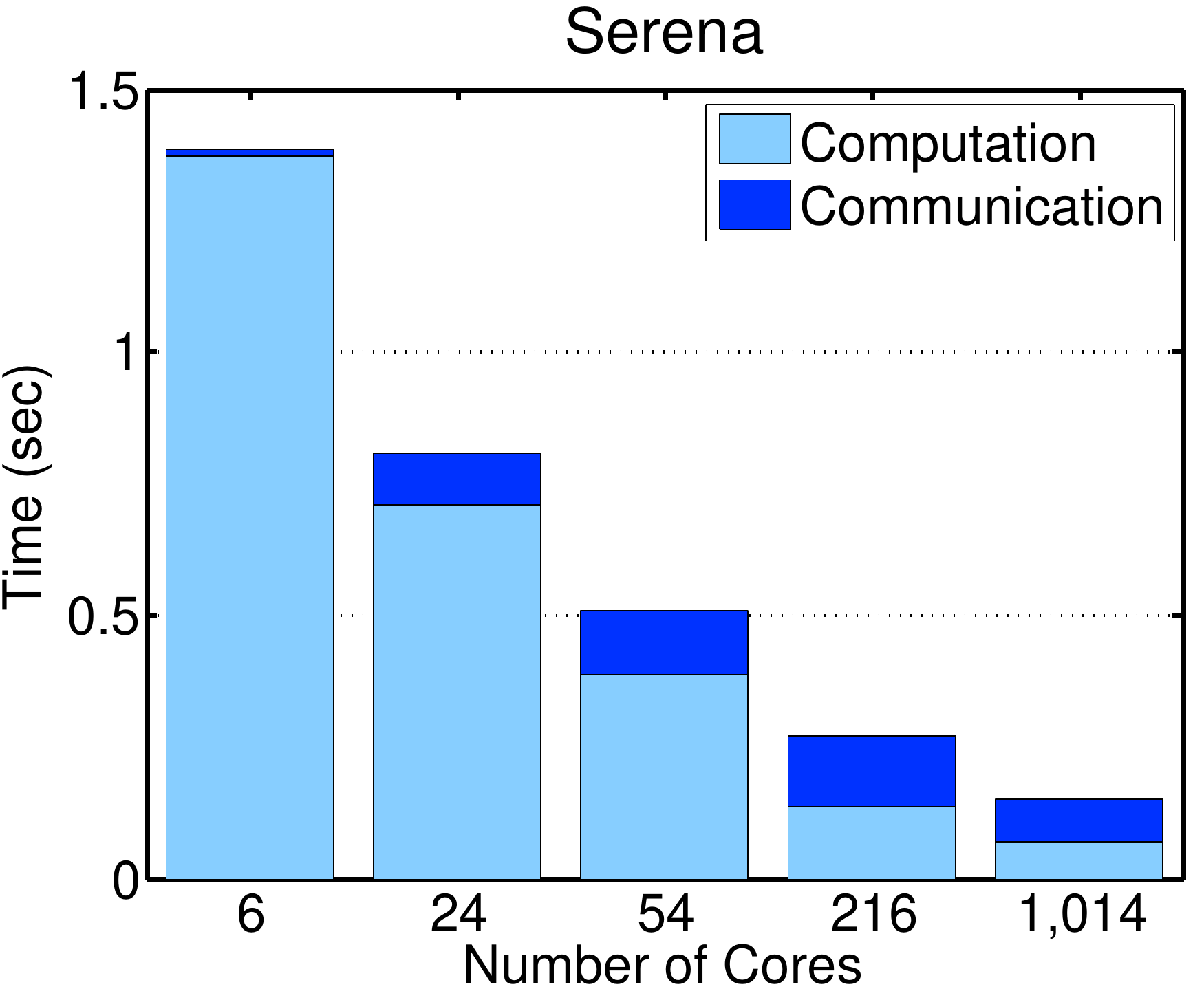}}
   \\
   \subfloat{\includegraphics[scale=.3]{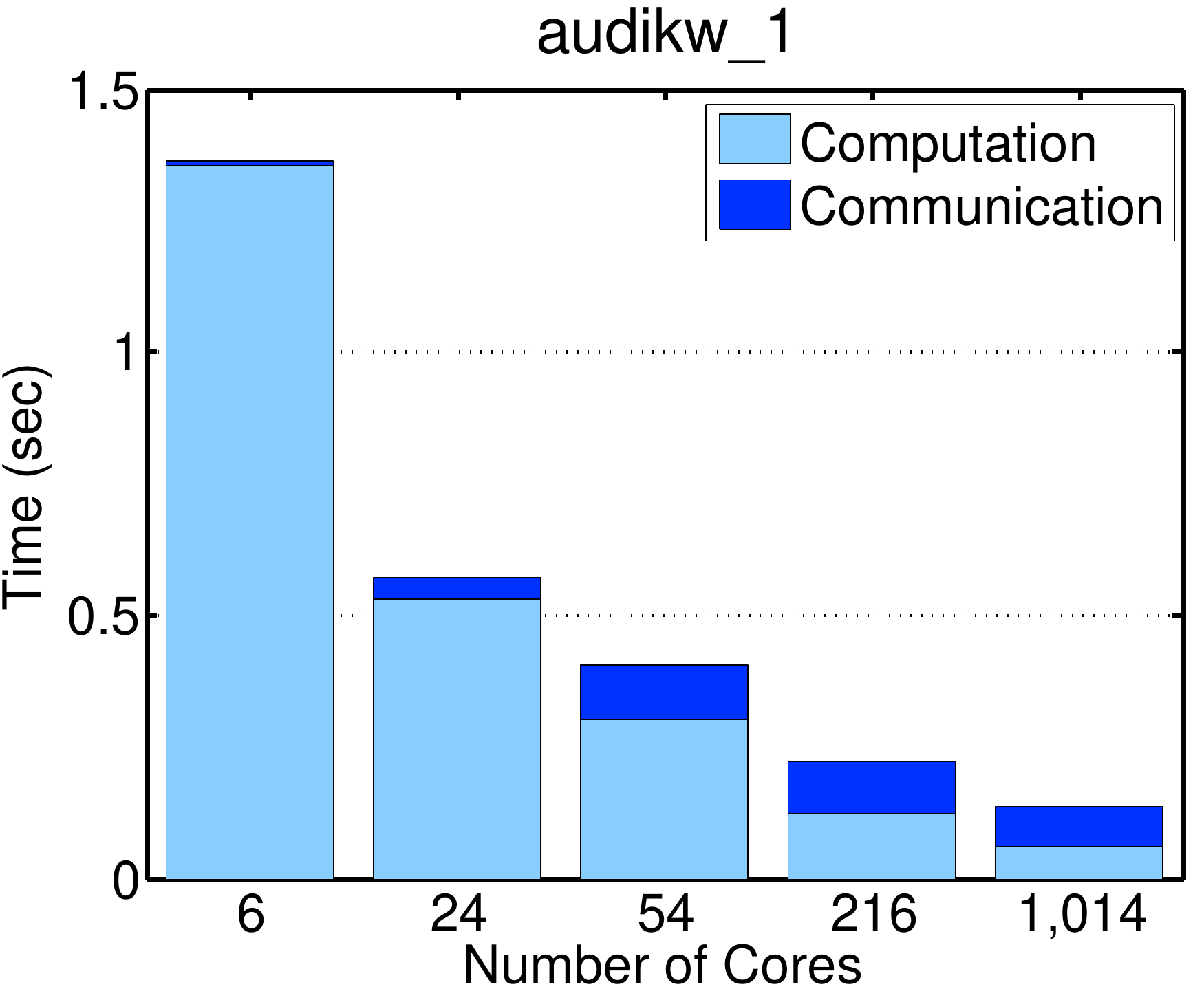}} 
   ~ ~
   \subfloat{\includegraphics[scale=.3]{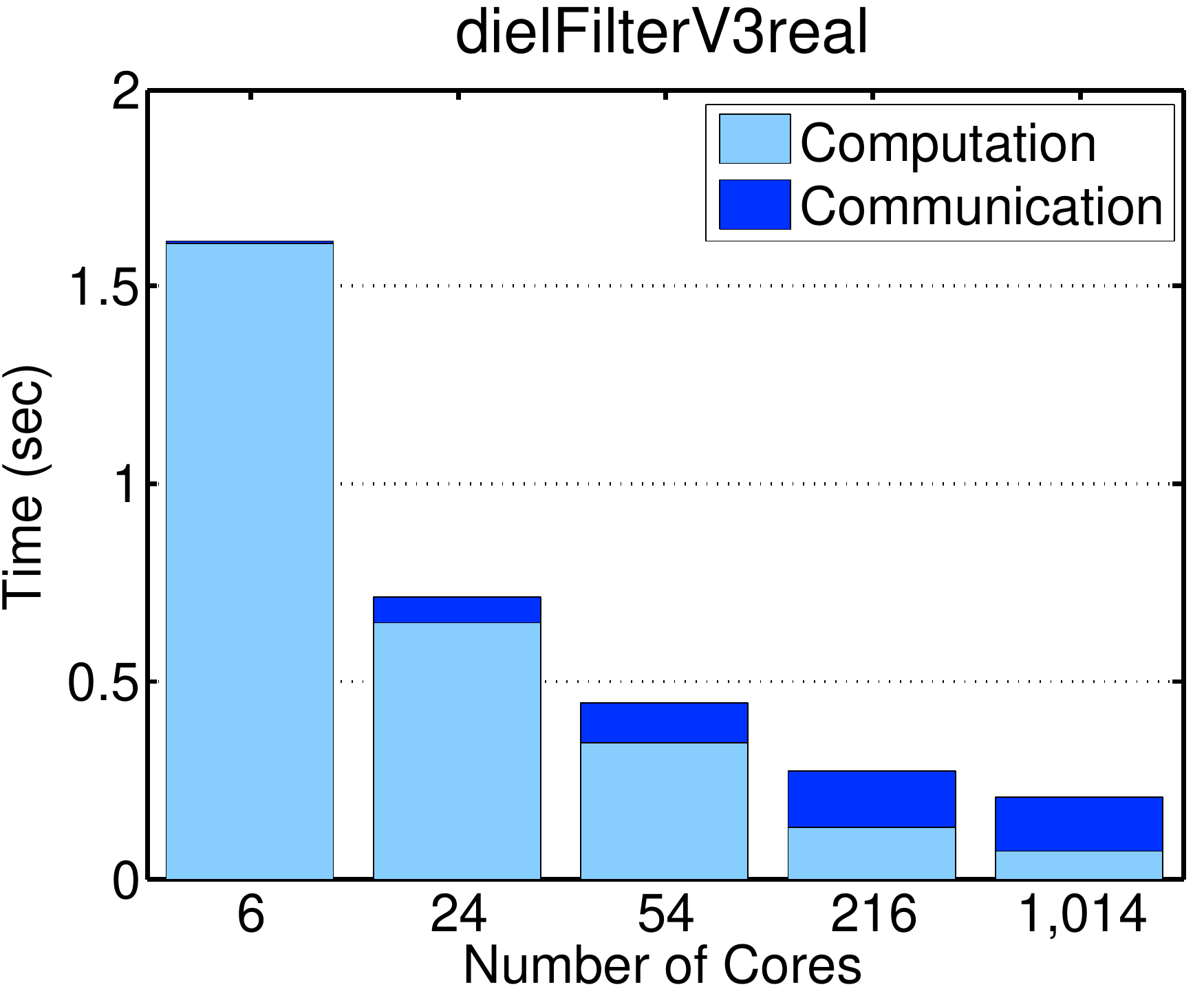}}
   ~ ~
   \subfloat{\includegraphics[scale=.3]{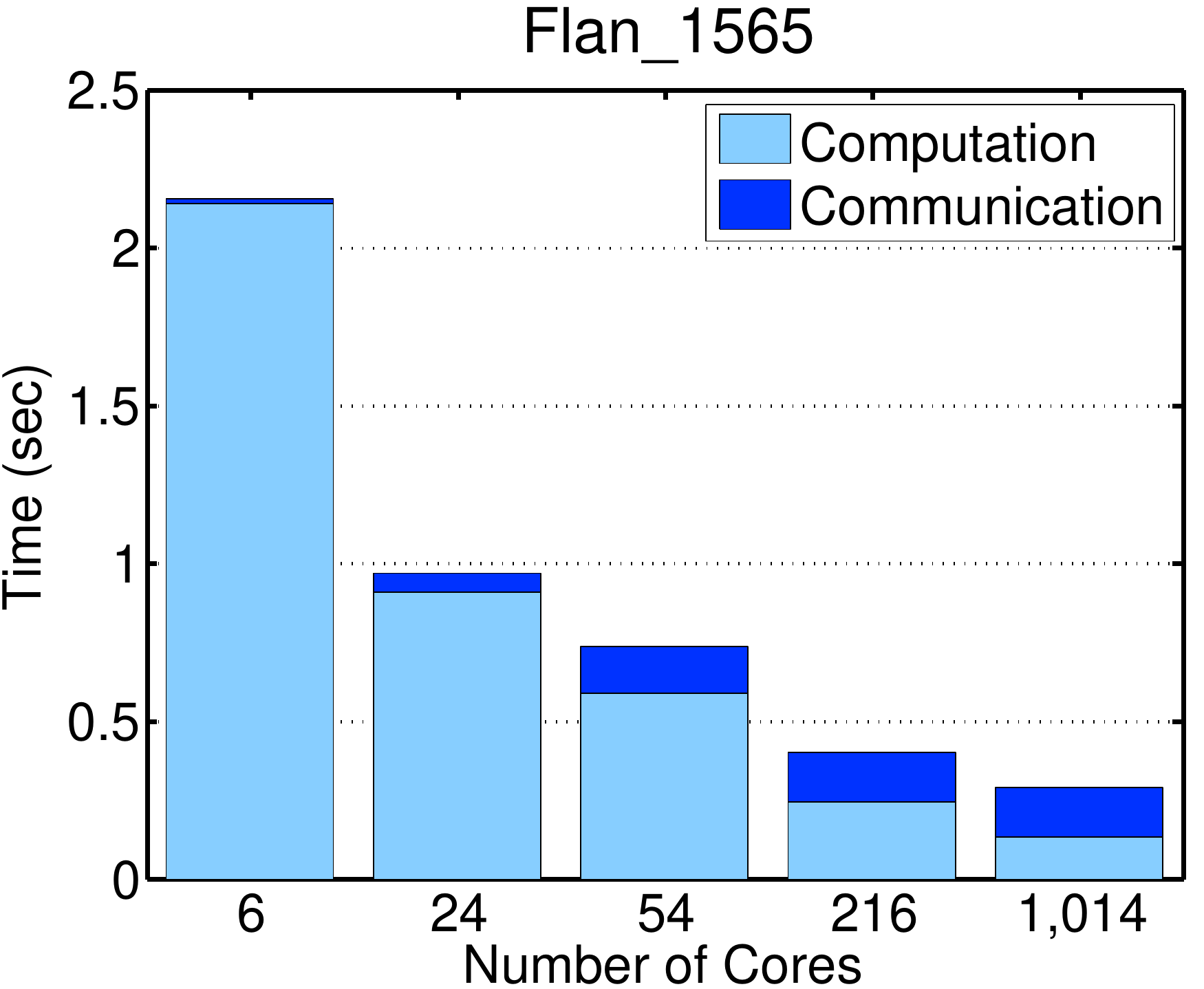}}
  \\

   \subfloat{\includegraphics[scale=.3]{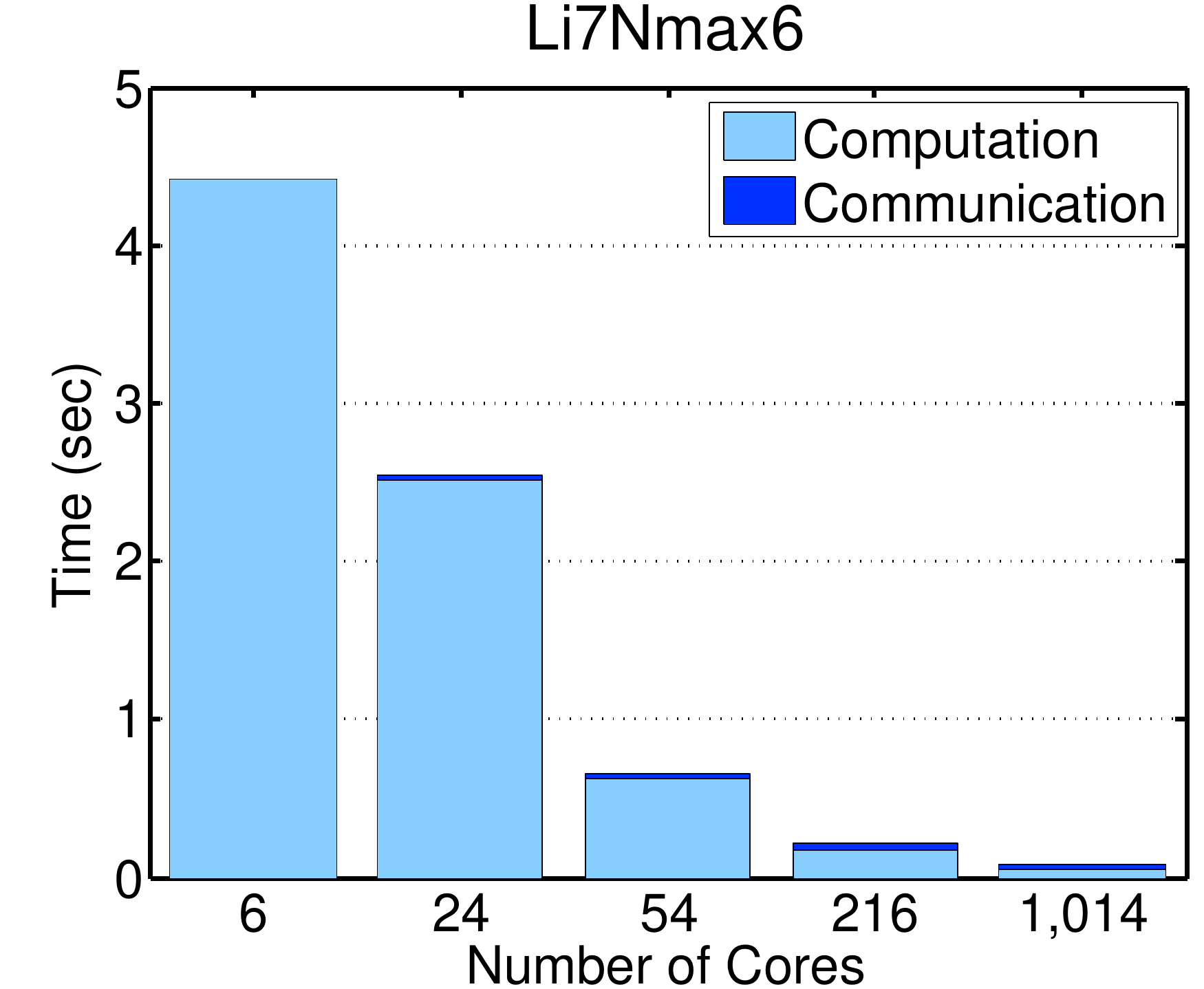}}
   ~~
   \subfloat{\includegraphics[scale=.3]{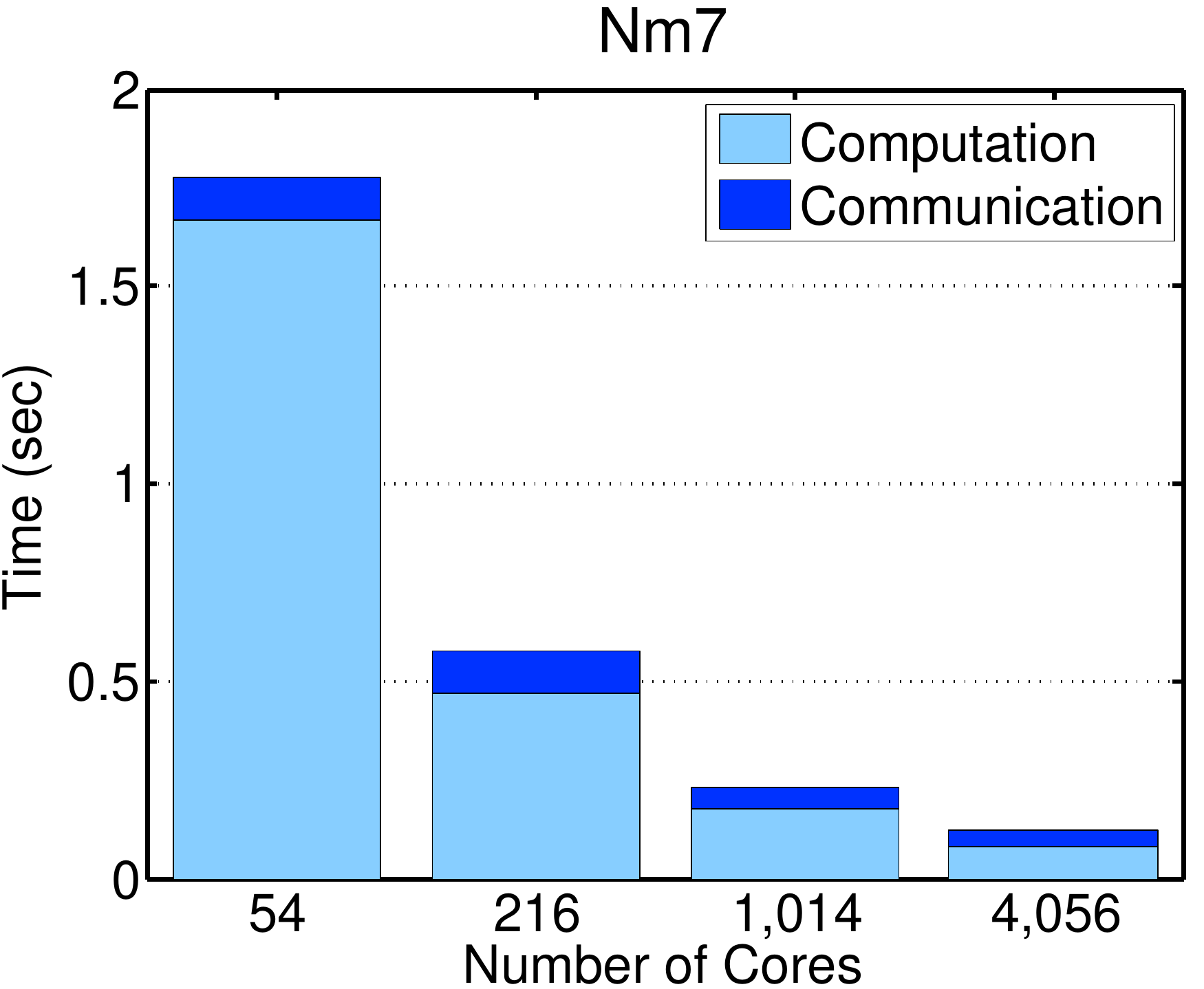}}
   ~ ~
   \subfloat{\includegraphics[scale=.3]{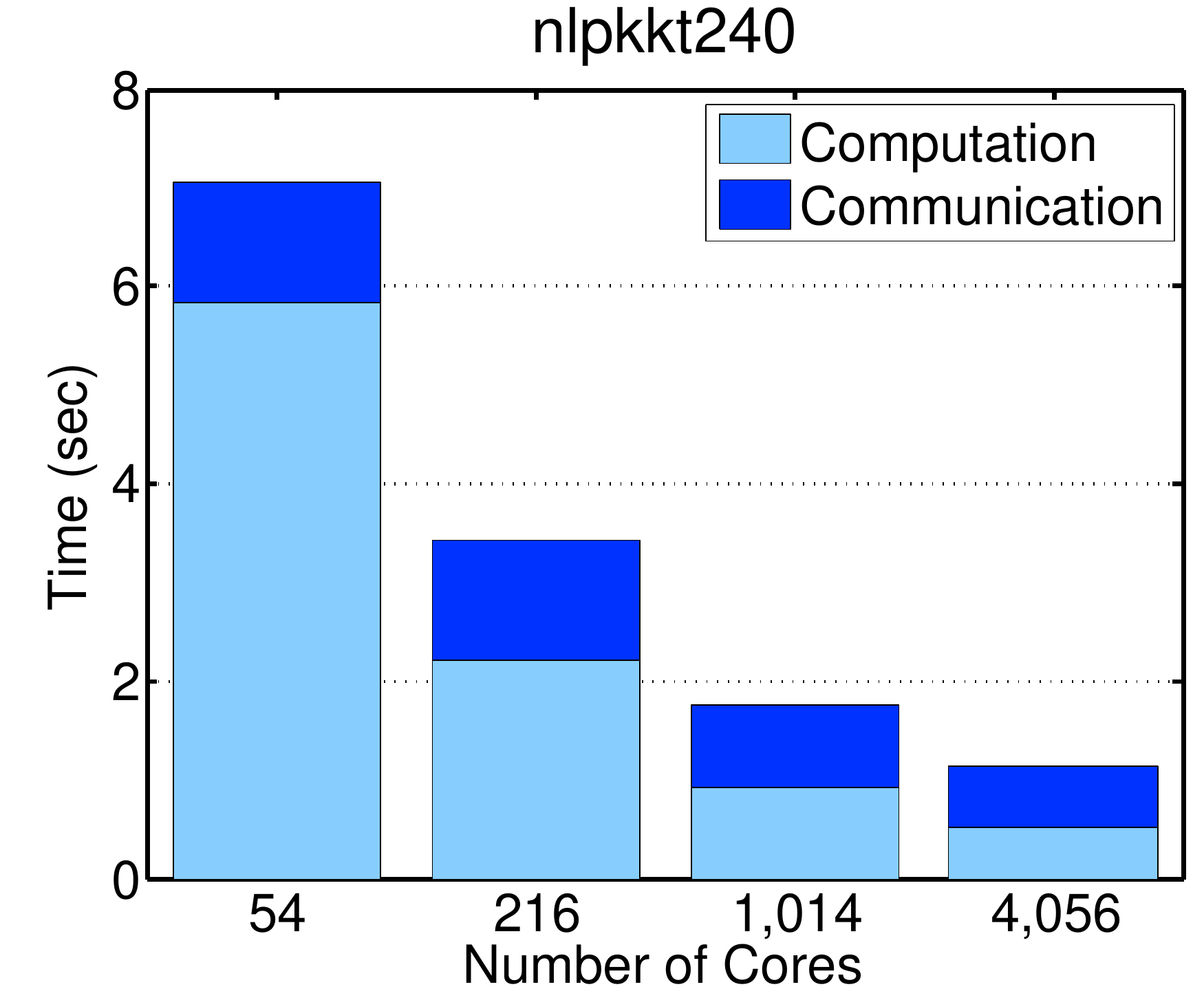}}

   \caption{Computation and communication time in all SpMSpV calls on different graphs on Edison. Six threads per MPI process are used.}
   \label{fig:comp_comm}
\end{figure*}

\begin{figure}[!t]
  \centering
   
  \includegraphics[scale=.33]{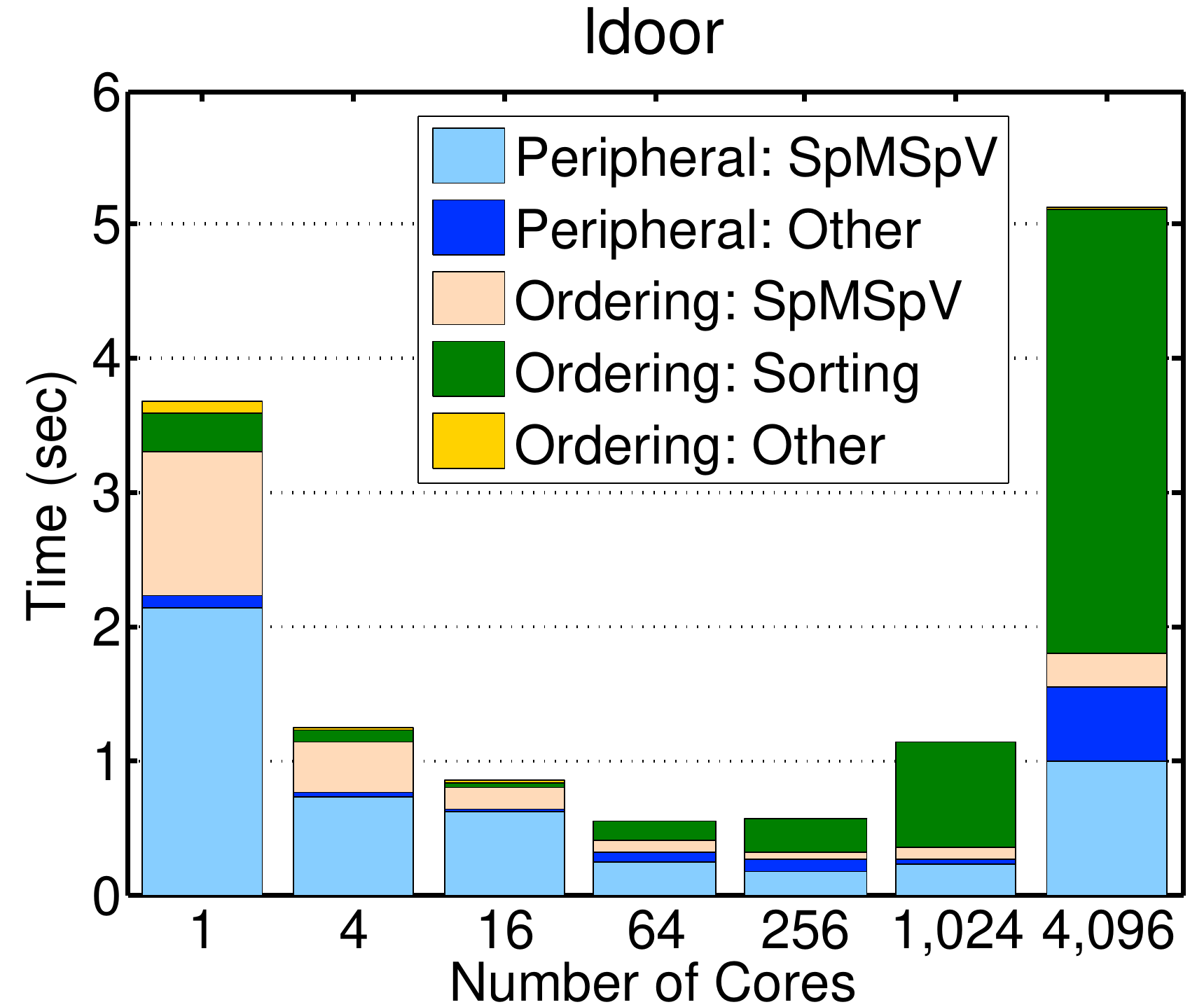}

\caption{Breakdown of total runtime when computing RCM ordering for \texttt{ldoor} using one thread per MPI process (flat MPI) on Edison.}
   \label{fig:ldoor_1t}
\end{figure}

\subsection{Distributed-memory performance}
We ran the distributed-memory RCM algorithm on up to 4096 cores of Edison. 
All performance results shown in this section used six threads per MPI process as it performed the best.
Figure~\ref{fig:timebreak_6t} shows the strong scaling of the distributed-memory RCM algorithm for nine graphs from Table~\ref{fig:testsuite}.
To better understand the performance, we break down the runtime into five parts at each concurrency where the height of a bar denotes the total runtime of identifying a pseudo-peripheral vertex and computing the RCM ordering.

Our distributed algorithm scales up to 1024 cores for all of the nine graphs as shown in Figure~\ref{fig:timebreak_6t}.
The RCM algorithm attains the best speedup of $38\times$ for \texttt{Li7Nmax6} and $27\times$ on \texttt{nd24k} on 1024 cores.
By contrast, it achieves $10\times$ and $13\times$ speedups for \texttt{ldoor} and \texttt{Flan\_1565}.
The sharp drop in parallel efficiency on these graphs is due to their relatively high diameters.
The level-synchronous nature of our BFS incurs high latency costs and decreases the amount of work per processor on high-diameter graphs.
Figure~\ref{fig:timebreak_6t} shows that \Call{SpMSpV}{} is usually the most expensive operation on lower concurrency.
However, \Call{SortPerm}{} starts to dominate on high concurrency because it performs an AllToAll among all processes, which has higher latency.
The size of the matrix also contributes to the scalability of the distributed-memory RCM algorithm.
For example, the largest two matrices in our test set (\texttt{Nm7} and nlpkkt240) continue to scale on more than 4K cores whereas smaller problems do not scale beyond 1K cores. 
This demonstrates that our algorithm can scale on even higher core count if larger matrices are given as inputs.

Since the distributed-memory \Call{SpMSpV}{} is the most expensive step of our RCM algorithm, we investigate its performance more closely.
Figure~\ref{fig:comp_comm} shows the breakdown of computation and communication time of the distributed \Call{SpMSpV}{} primitive. 
We observe that on lower concurrency, \Call{SpMSpV}{} is dominated by its computation, as expected.
The communication time of \Call{SpMSpV}{} starts to dominate the computation time on higher concurrency, and the crossover point where the communication becomes more expensive than computation depends on the properties of the matrices.
Graphs with higher diameters have higher communication overhead than graphs with low diameters. For example, \texttt{ldoor} has one of the highest pseudo-diameters among problems we considered.
Hence, the RCM algorithm on  \texttt{ldoor} becomes communication bound on 1K cores. 
By contrast, other lower diameter graphs with similar sizes continue to compute bound and scale beyond 1K cores on Edison.


We have also experimented with a flat MPI approach. This non-threaded RCM implementation had higher communication overhead and ran slower than the multithreaded implementation, especially on higher concurrencies. For example, the flat MPI implementation took $5\times$ longer to compute RCM on the \texttt{ldoor} matrix using 4096 cores of Edison as shown in Figure~\ref{fig:ldoor_1t}.


\section{Conclusion and Future Work}
\label{sec:conclusion}
We have introduced the first distributed-memory implementation of the RCM algorithm.  In particular, we have described how the RCM algorithm can be reformulated using the sparse matrix / graph duality, so that the parallel implementation can be accomplished
using a small set of parallel primitives that have been highly optimized.
Our experiments have shown that the distributed-memory implementation of the RCM algorithm presented in this work scales well up to 1024 processors for smaller matrices, and even to 4096 cores for the largest problems we have evaluated.

We have shown that this new approach is overall faster than the traditional approach that
gathers the matrix structure on a single node, followed by the application of a serial or a
multithreaded implementation of the RCM algorithm, and then redistributing the permuted matrix. More importantly, our approach removes the memory bottleneck that may be caused by having to store the entire graph with a single node.

A scalable implementation of the RCM algorithm is of prime importance for many iterative solvers that benefit from
reordering the matrix. We have assessed this on a sample
problem using the conjugate gradient method with block Jacobi preconditioner available in PETSc.

Immediate future work involves finding alternatives to sorting (i.e. global sorting at the end, or not sorting at all and sacrifice some quality).
Longer term work would investigate alternative BFS formulations that are not level-synchronous, as the existing approach has trouble scaling to high-diameter graphs.

\section*{Acknowledgments}
We thank Hari Sundar for sharing the source code of HykSort.
This work is supported by the Applied Mathematics and the SciDAC Programs of the DOE Office of Advanced Scientific Computing Research under contract number DE-AC02-05\-CH\-11231.
We used resources of the NERSC supported by the Office of Science of the DOE under Contract No. DE-AC02-05CH11231. 

\vspace{-8pt}

\bibliographystyle{IEEEtran}

\bibliography{RCM}

\begin{thebibliography}{10}
\providecommand{\url}[1]{#1}
\csname url@samestyle\endcsname
\providecommand{\newblock}{\relax}
\providecommand{\bibinfo}[2]{#2}
\providecommand{\BIBentrySTDinterwordspacing}{\spaceskip=0pt\relax}
\providecommand{\BIBentryALTinterwordstretchfactor}{4}
\providecommand{\BIBentryALTinterwordspacing}{\spaceskip=\fontdimen2\font plus
\BIBentryALTinterwordstretchfactor\fontdimen3\font minus
  \fontdimen4\font\relax}
\providecommand{\BIBforeignlanguage}[2]{{%
\expandafter\ifx\csname l@#1\endcsname\relax
\typeout{** WARNING: IEEEtran.bst: No hyphenation pattern has been}%
\typeout{** loaded for the language `#1'. Using the pattern for}%
\typeout{** the default language instead.}%
\else
\language=\csname l@#1\endcsname
\fi
#2}}
\providecommand{\BIBdecl}{\relax}
\BIBdecl

\bibitem{duff1989use}
I.~S. Duff, J.~K. Reid, and J.~A. Scott, ``The use of profile reduction
  algorithms with a frontal code,'' \emph{International Journal for Numerical
  Methods in Engineering}, vol.~28, no.~11, pp. 2555--2568, 1989.

\bibitem{duff1989effect}
I.~S. Duff and G.~A. Meurant, ``The effect of ordering on preconditioned
  conjugate gradients,'' \emph{BIT Numerical Mathematics}, vol.~29, no.~4, pp.
  635--657, 1989.

\bibitem{papadimitriou1976np}
C.~H. Papadimitriou, ``The {NP}-completeness of the bandwidth minimization
  problem,'' \emph{Computing}, vol.~16, no.~3, pp. 263--270, 1976.

\bibitem{cuthill1969reducing}
E.~Cuthill and J.~McKee, ``Reducing the bandwidth of sparse symmetric
  matrices,'' in \emph{Proc.\ of 24th national conference}.\hskip 1em plus
  0.5em minus 0.4em\relax ACM, 1969, pp. 157--172.

\bibitem{GeorgeLiu81}
A.~George and J.~W.-H. Liu, \emph{Computer Solution of Large Sparse Positive
  Definite Systems}.\hskip 1em plus 0.5em minus 0.4em\relax Englewood Cliffs,
  New Jersey: Prentice-Hall Inc., 1981.

\bibitem{sloan1986algorithm}
S.~Sloan, ``An algorithm for profile and wavefront reduction of sparse
  matrices,'' \emph{International Journal for Numerical Methods in
  Engineering}, vol.~23, no.~2, pp. 239--251, 1986.

\bibitem{petsc-user-ref}
\BIBentryALTinterwordspacing
S.~Balay, S.~Abhyankar, M.~F. Adams, J.~Brown, P.~Brune, K.~Buschelman,
  L.~Dalcin, V.~Eijkhout, W.~D. Gropp, D.~Kaushik, M.~G. Knepley, L.~C.
  McInnes, K.~Rupp, B.~F. Smith, S.~Zampini, H.~Zhang, and H.~Zhang, ``{PETS}c
  users manual,'' Argonne National Laboratory, Tech. Rep. ANL-95/11 - Revision
  3.7, 2016. [Online]. Available: \url{http://www.mcs.anl.gov/petsc}
\BIBentrySTDinterwordspacing

\bibitem{spmp-rcm}
K.~I. Karantasis, A.~Lenharth, D.~Nguyen, M.~J. Garzaran, and K.~Pingali,
  ``Parallelization of reordering algorithms for bandwidth and wavefront
  reduction,'' in \emph{International Conference for High Performance
  Computing, Networking, Storage and Analysis (SC'14)}.\hskip 1em plus 0.5em
  minus 0.4em\relax IEEE, 2014, pp. 921--932.

\bibitem{george1971}
J.~A. George, ``Computer implementation of the finite element method,'' Ph.D.
  dissertation, Stanford, CA, USA, 1971.

\bibitem{berge}
B.~Claude, \emph{The Theory of Graphs and its Applications}.\hskip 1em plus
  0.5em minus 0.4em\relax John Wiley \& Sons, 1962.

\bibitem{arany}
I.~Arany, L.~Szoda, and W.~Smyth, ``An improved method for reducing the
  bandwidth of sparse symmetric matrices.'' in \emph{IFIP Congress (2)}, 1971,
  pp. 1246--1250.

\bibitem{george1979}
A.~George and J.~W.~H. Liu, ``An implementation of a pseudoperipheral node
  finder,'' \emph{ACM Transactions on Mathematical Software (TOMS)}, vol.~5,
  no.~3, pp. 284--295, Sep. 1979.

\bibitem{gibbs1976algorithm}
N.~E. Gibbs, W.~G. Poole, Jr, and P.~K. Stockmeyer, ``An algorithm for reducing
  the bandwidth and profile of a sparse matrix,'' \emph{SIAM Journal on
  Numerical Analysis (SINUM)}, vol.~13, no.~2, pp. 236--250, 1976.

\bibitem{bfs:11}
A.~Bulu{\c{c}} and K.~Madduri, ``Parallel breadth-first search on distributed
  memory systems,'' in \emph{International Conference for High Performance
  Computing, Networking, Storage and Analysis (SC'11)}.\hskip 1em plus 0.5em
  minus 0.4em\relax ACM, 2011, pp. 65:1--65:12.

\bibitem{checconi2012breaking}
F.~Checconi, F.~Petrini, J.~Willcock, A.~Lumsdaine, A.~R. Choudhury, and
  Y.~Sabharwal, ``Breaking the speed and scalability barriers for graph
  exploration on distributed-memory machines,'' in \emph{International
  Conference for High Performance Computing, Networking, Storage and Analysis
  (SC'12)}.\hskip 1em plus 0.5em minus 0.4em\relax IEEE, 2012, pp. 1--12.

\bibitem{ashcraft1987progress}
C.~C. Ashcraft, R.~G. Grimes, J.~G. Lewis, B.~W. Peyton, H.~D. Simon, and P.~E.
  Bj{\o}rstad, ``Progress in sparse matrix methods for large linear systems on
  vector supercomputers,'' \emph{International Journal of High-Performance
  Computing Applications {(IJHPCA)}}, vol.~1, no.~4, pp. 10--30, 1987.

\bibitem{bulucc2011combinatorial}
A.~Bulu{\c{c}} and J.~R. Gilbert, ``The {C}ombinatorial {BLAS}: Design,
  implementation, and applications,'' \emph{International Journal of
  High-Performance Computing Applications {(IJHPCA)}}, vol.~25, no.~4, 2011.

\bibitem{matchingipdps16}
A.~Azad and A.~Bulu\c{c}, ``Distributed-memory algorithms for maximum
  cardinality matching in bipartite graphs,'' in \emph{IEEE International
  Parallel \& Distributed Processing Symposium (IPDPS)}, 2016.

\bibitem{bruck1997efficient}
J.~Bruck, C.-T. Ho, S.~Kipnis, E.~Upfal, and D.~Weathersby, ``Efficient
  algorithms for all-to-all communications in multiport message-passing
  systems,'' \emph{IEEE Transactions on Parallel and Distributed Systems
  (TPDS)}, vol.~8, no.~11, pp. 1143--1156, 1997.

\bibitem{sundar2013hyksort}
H.~Sundar, D.~Malhotra, and G.~Biros, ``{HykSort}: a new variant of hypercube
  quicksort on distributed memory architectures,'' in \emph{International
  Conference on Supercomputing (ICS)}.\hskip 1em plus 0.5em minus 0.4em\relax
  ACM, 2013, pp. 293--302.

\bibitem{ufget}
T.~A. Davis and Y.~Hu, ``The university of florida sparse matrix collection,''
  \emph{ACM Transactions on Mathematical Software (TOMS)}, vol.~38, no.~1,
  p.~1, 2011.

\bibitem{ipdps14}
H.~M. Aktulga, A.~Bulu\c{c}, S.~Williams, and C.~Yang, ``Optimizing sparse
  matrix-multiple vectors multiplication for nuclear configuration interaction
  calculations,'' in \emph{IEEE International Parallel \& Distributed
  Processing Symposium (IPDPS)}.\hskip 1em plus 0.5em minus 0.4em\relax IEEE
  Computer Society, 2014.

\bibitem{spmp}
\BIBentryALTinterwordspacing
J.~{Park et al.}, ``{SpMP}: {S}parse {M}atrix {P}re-processing.'' [Online].
  Available: \url{https://github.com/jspark1105/SpMP}
\BIBentrySTDinterwordspacing

\bibitem{spmp-optim}
J.~Chhugani, N.~Satish, C.~Kim, J.~Sewall, and P.~Dubey, ``Fast and efficient
  graph traversal algorithm for {CPU}s: Maximizing single-node efficiency,'' in
  \emph{IEEE International Parallel \& Distributed Processing Symposium
  (IPDPS)}.\hskip 1em plus 0.5em minus 0.4em\relax IEEE, 2012, pp. 378--389.

\end{thebibliography}

\end{document}